\documentclass[useAMS,usenatbib,usegraphicx]{mn2e}
\usepackage{amssymb,amsmath, color, url,soul}

\newcommand{\msano}{{\rm M}_\odot ~{\rm yr}^{-1}}
\newcommand{\alf}{Alfv\'en}
\newcommand{\jdot}{\dot{J}}
\newcommand{\mdot}{\dot{M}}

\title{On the environment surrounding close-in exoplanets}
\author[A.~A.~Vidotto et al.]{A.~A.~Vidotto$^{1}$\thanks{E-mail: Aline.Vidotto@unige.ch}, {R.~Fares}$^{2}$, {M.~Jardine}$^{2}$, {C.~Moutou}$^{3,4}$, {J.-F.~Donati}$^{5}$
\\ 
$^{1}$Observatoire de Gen\`eve, Universit\'e de Gen\`eve, Chemin des Maillettes 51, Versoix, CH-1290, Switzerland\\
$^{2}$SUPA, School of Physics and Astronomy, University of St Andrews, North Haugh, St Andrews, KY16 9SS, UK\\
$^{3}$Canada-France-Hawaii Telescope Corporation, CNRS, 65-1238 Mamalahoa Hwy, Kamuela HI 96743, USA\\
$^{4}$Aix Marseille Universit\'e, CNRS, LAM (Laboratoire d'Astrophysique de Marseille) UMR 7326, 13388, Marseille, France\\
$^{5}$LATT-UMR 5572, CNRS \& Univ. P. Sabatier, 14 Av.~E.~Belin, Toulouse, F-31400, France
}
\begin{document}

\date{Accepted . Received ; in original form}

\pagerange{\pageref{firstpage}--\pageref{lastpage}} \pubyear{2014}

\maketitle

\label{firstpage}

\begin{abstract}
Exoplanets in extremely close-in orbits are immersed in a local interplanetary medium (i.e., the stellar wind) much denser than the local conditions encountered around the solar system planets. The environment surrounding these exoplanets also differs in terms of dynamics (slower stellar winds, but higher Keplerian velocities) and ambient magnetic fields (likely higher for host stars more active than the Sun). Here, we quantitatively investigate the nature of the interplanetary media surrounding the hot Jupiters HD46375b, HD73256b, HD102195b, HD130322b, HD179949b. We simulate the three-dimensional winds of their host stars, in which we directly incorporate their observed surface magnetic fields. With that, we derive mass-loss rates ($1.9$ to $8.0 \times 10^{-13}~\msano$) and the wind properties at the position of the hot-Jupiters' orbits (temperature, velocity, magnetic field intensity and pressure). We show that these exoplanets' orbits are super-magnetosonic, indicating that bow shocks are formed surrounding these planets. Assuming planetary magnetic fields similar to Jupiter's, we estimate planetary magnetospheric sizes of $4.1$ to $5.6$ planetary radii.
We also derive the exoplanetary radio emission released in the dissipation of the stellar wind energy. We find radio fluxes ranging from $0.02$ to $0.13$~mJy, which are challenging to be observed with present-day technology, but could be detectable with future higher sensitivity arrays (e.g., SKA). Radio emission from systems having closer hot-Jupiters, such as from $\tau$~Boo~b or HD~189733b, or from nearby planetary systems orbiting young stars, are likely to have higher radio fluxes, presenting better prospects for detecting exoplanetary radio emission.
\end{abstract}

\section{INTRODUCTION}
The interplanetary medium that surrounds exoplanets is filled by stellar wind particles and the embedded stellar magnetic field. Due to detection biases, the large majority of exoplanets found so far are orbiting cool stars at the main-sequence phase. Although the winds of these stars have proven quite challenging to observe \citep{1992ApJ...397..225M,2001ApJ...546L..57W,2005ApJ...628L.143W}, the interaction between exoplanets and their surrounding medium (i.e., the host star's wind) may give rise to observable signatures, such as planetary radio emission \citep{2007P&SS...55..598Z}, enhancement of stellar activity \citep{2000ApJ...533L.151C,2003ApJ...597.1092S,2005ApJ...622.1075S}, bow-shock formation \citep{2010ApJ...722L.168V,2013MNRAS.436.2179L,2013ApJ...764...19B}, charge-exchange between stellar wind protons and planetary neutral hydrogen \citep{2008Natur.451..970H,2010ApJ...709..670E,2013A&A...557A.124B,2014A&A...562A.116K} and formation of comet-like tail structures \citep{2011Icar..211....1M,2012ApJ...752....1R,2013A&A...557A..72B,2014MNRAS.438.1654V}, all of which can provide invaluable insights into the system, such as the intensity of the planetary magnetic field, velocity and temperature of the local stellar wind, etc.

By studying stellar winds, we are able to make quantitative predictions about the interplanetary medium. A significant improvement on our understanding of the interaction between a planet and the wind of its host star has been achieved in the past decade. Traditionally, these works have been based on simplified treatments of the winds \citep[e.g.][]{2004ApJ...602L..53I,2005A&A...434.1191P,2005A&A...437..717G,2005MNRAS.356.1053S,2008MNRAS.389.1233L, 2011MNRAS.411L..46V, 2014A&A...570A..99S,  2014ApJ...795...86S}. For example, simplified wind approaches might assume an isothermal wind structure, or that stars are non-rotating and/or non-magnetised bodies, among others. However, stellar winds are three-dimensional (3D) in nature, where complex interactions of a rotating, magnetised plasma take place. In view of that, more recently, new generations of 3D, magnetohydrodynamical (MHD) models have started to be employed in the studies of interactions between stars/winds and their planets \citep[e.g.,][]{2009ApJ...703.1734V,2009ApJ...699..441V,2010ApJ...720.1262V,2011MNRAS.412..351V,2012MNRAS.423.3285V,2014MNRAS.438.1162V, 2009ApJ...704L..85C, 2014ApJ...790...57C, 2013MNRAS.436.2179L}. 

The advantage of using simplified treatments of the wind is that these works rely on analytical and low-dimensional (1D, 2D) numerical studies, which are significantly faster and do not demand extensive computational resources as 3D models do. Because of that, 1D works can investigate a much wider range of stellar wind parameters \citep[e.g.][]{2011ApJ...741...54C,2014A&A...570A..99S} than more complex, computationally-expensive 3D models can. The disadvantage, on the other hand, is that the simplified models can not capture the 3D structure of stellar winds. Combined with modern techniques to reconstruct stellar magnetic fields, some 3D models are able to provide a more realistic account of the stellar magnetic field topology embedded in the wind, recognised to be essential to interpret and predict signatures of star-planet interactions \citep{2006MNRAS.367L...1M,2010MNRAS.406..409F,2011MNRAS.414.1573V,2012A&A...544A..23L, 2013MNRAS.436.2179L}. 

\subsection{Interactions with magnetised planets}\label{sec.introBp}
As the wind outflows from the star, it interacts with any planet encountered on its way. If these planets are magnetised, their magnetic fields can act as shields, which prevent stellar wind particles from reaching all the way down to the surface or atmosphere of these objects \citep[e.g.][]{2007AsBio...7..167K,2007AsBio...7..185L,2013Icar..226.1447D,2014A&A...562A.116K}. This is in particular the case of the Earth and, more generally, of planets with dipolar field configurations. For these objects, the solar and stellar winds are deflected around the magnetospheric cavity,  potentially helping the planet to retain its atmosphere. However, atmospheric escape can still occur at high magnetic latitudes through polar flows, as is the case of the Earth (e.g., \citealt{2001Sci...291.1939S,2007RvGeo..45.3002M}) and predicted for exoplanets (\citealt{2014MNRAS.444.3761O}; see also Section~\ref{sec.polarflows}). Part of this planetary outflow can return from the magnetosphere back into atmospheric regions of low-magnetic latitudes, reducing the total net loss rate of atmospheric escape, as suggested for the Earth scenario \citep{2001Sci...291.1939S}. The detailed process of atmospheric dynamics and escape is certainly complex and not examined here.

In the present work, only magnetised exoplanets are considered. This means that the cross-section of the  `obstacle' is not that of the planet itself, but rather takes into account the magnetospheric size of the planet. The magnetospheric size of the planet depends both on the the characteristics of the local environment surrounding the planet (interplanetary density, velocity, magnetic field, temperature) and on its own magnetic field. On the theoretical side, some models suggest that the strength of the planetary magnetic field is dependent on the rotation rate of the planet \citep{1999JGR...10414025F}. In this situation, close-in planets that are tidally locked could have a reduced magnetic moment \citep{2004A&A...425..753G}. Other models advocate that the planetary magnetic field is related to the energy flux coming from the planetary core and does not depend on the rotation rate of the planet \citep{2009Natur.457..167C}. Recent studies indicate that the planetary field strength is independent of rotation rate, which instead plays a role in the geometry of the generated magnetic field \citep{2012Icar..217...88Z}.

Although planetary magnetism has been observed in several solar system planets, such as in Earth and the giant planets, the presence of exoplanetary magnetic fields are much more elusive. \citet{2010ApJ...722L.168V} suggested that the close-in giant planet WASP-12b hosts a bow-shock that surrounds its magnetosphere at a distance of about $4$ -- $5$ planetary radii. Their suggestion was motivated by transit observations of the close-in giant planet WASP-12b by \citet{2010ApJ...714L.222F}, who, based on space-borne spectroscopic observations in the near-UV, showed that the transit lightcurve of WASP-12b presents both an early ingress when compared to its optical transit, as well as excess absorption during the transit \citep[see also][]{2012ApJ...760...79H}. \citet{2010ApJ...722L.168V} attributed this signature to an absorption of the material in the bow shock (see also \citealt{2011MNRAS.416L..41L}). If confirmed, this technique should provide a useful tool for determining planetary magnetic field intensities for hot-Jupiter transiting systems. In the case of WASP-12b, \citet{2010ApJ...722L.168V} derived an upper limit of $24$~G for the planetary field. \citet{2011MNRAS.411L..46V} later proposed other targets with good prospects to hosting observable early-ingresses. Unfortunately, the near-UV ($254-258$nm) early-ingress signature of WASP-12b observed with (expensive) space-based spectroscopic observations \citep{2010ApJ...714L.222F, 2012ApJ...760...79H} does not seem to be observable with ground-based, broad-band photometry  in the wavelength range $\sim 340 - 540$nm \citep{2012ApJ...760...79H}, and neither in the range of $303 - 417$nm (Turner, private comm.; for other transiting exoplanets see \citealt{2013MNRAS.428..678T,2014NewA...27..102P}). Observations from \citet{2010ApJ...714L.222F} indicate that the material surrounding WASP-12b absorbs at certain resonance lines in the near-UV (in particular in MgII lines). The lack of absorption from broad-band photometric observations of WASP-12b possibly indicates that either the material is not absorbing at the observed photometric wavelengths ($\sim 303 - 540$nm), or that the absorption occurs only at some specific spectral lines, but gets diluted over the much wider spectral region.

Another hint that close-in planets may also harbour intrinsic magnetic fields, similar to the Earth and the giant planets of the Solar System, was found by \citet{2003ApJ...597.1092S,2005ApJ...622.1075S,2008ApJ...676..628S}, who observed modulations of chromospheric spectral lines in phase with orbital periods on a few systems. Such modulations were interpreted as induced activity on the stellar surface due to magnetic interactions between star and planet. \citet{2008ApJ...676..628S} showed that there exists a correlation between the night-to-night stellar activity variation with the ratio between the planetary mass to orbital period, used as a proxy for the magnetic moment of a tidally-locked planet. Although unfortunately this correlation does not provide the intensity of the planetary magnetic field, it offers a way to measure the relative field strength among the different exoplanets in their sample. Therefore, once magnetism is assessed for one of their targets (by a different method), the magnetic field strength of their remaining targets could be derived.

These two suggestions (early ingress and activity enhancement), however, cannot be used as conclusive evidence of the presence of planetary magnetic fields, as alternative, non-magnetic explanations for the observations exist  \citep{2006A&A...460..317P,2010ApJ...721..923L, 2013ApJ...764...19B, vidotto_springer}. A conclusive way to probe the presence of exoplanetary magnetic fields could be achieved by the detection of radio emission from the planet. The stellar wind that impacts on the planet produces energetic particles that are captured by the planet's magnetic field, emitting cyclotron radiation at radio wavelengths. This emission depends on the planet's magnetic field intensity and on the stellar wind power: it implies that the stronger is the stellar wind, the more luminous is the planet. As such radio emission is observed in the Solar System \citep{2007P&SS...55..598Z}, there are expectations that close-in exoplanets will exhibit comparable radiation (see  \citealt{2012MNRAS.427L..75N} for the case of planets that are not necessarily close-in). In particular, hot-Jupiters are expected to be much more luminous than the most luminous planet in our solar System, Jupiter \citep[e.g.,][]{1999JGR...10414025F,2005A&A...437..717G,2007P&SS...55..598Z,2008A&A...490..843J,2010ApJ...720.1262V}. This is because hot-Jupiters are located much closer to their stars, interacting with portions of the host-star's wind that has larger kinetic and magnetic energies available to power planetary radio emission. So far, radio signatures of close-in exoplanets have not yet been detected \citep[e.g.][]{2000ApJ...545.1058B,2004ApJ...612..511L,2009MNRAS.395..335S,2013ApJ...762...34H} and one possible reason for that may be due to the lack of instrumental sensitivity in the appropriate frequency range of the observations \citep{2000ApJ...545.1058B}. This picture, however, might be changing, as possible hints of exoplanetary radio emission have recently been reported \citep{2013A&A...552A..65L,2014A&A...562A.108S}. 

The theoretical estimates of the radio flux emitted by extrasolar planets carry a large uncertainty due to the fact that the stellar wind properties are poorly constrained. In this work, we model the 3-D structure of the stellar wind of a sample of $5$ planet-hosting stars, whose 3-D winds have not yet been studied to date. We investigate the nature of the interplanetary media of these exoplanetary systems and how different they are from the environment surrounding our own Solar System planets. The stars used in this study, described in Section~\ref{sec.sample}, have had their surface magnetic field recently reconstructed by means of tomographic techniques \citep{2012MNRAS.423.1006F,2013MNRAS.435.1451F}. These surface fields are used as boundary conditions for our data-driven simulations of stellar winds. Our model is described in Section~\ref{sec.model}. The derived global characteristics of the stellar winds are presented in Section~\ref{sec.results} and the properties of the local environment surrounding the exoplanets in our sample are described in Section~\ref{sec.planets}. We then use these computed quantities to calculate the strengths of the interactions between the stellar wind and the planetary system, making it possible to quantitatively predict planetary radio emission and bow shock formation. Our discussion is shown in Section \ref{sec.discussion}
and summary and conclusions are presented in Section~\ref{sec.conclusions}.

\section{The sample of stars}\label{sec.sample}
\begin{figure*}
\includegraphics[width=58mm]{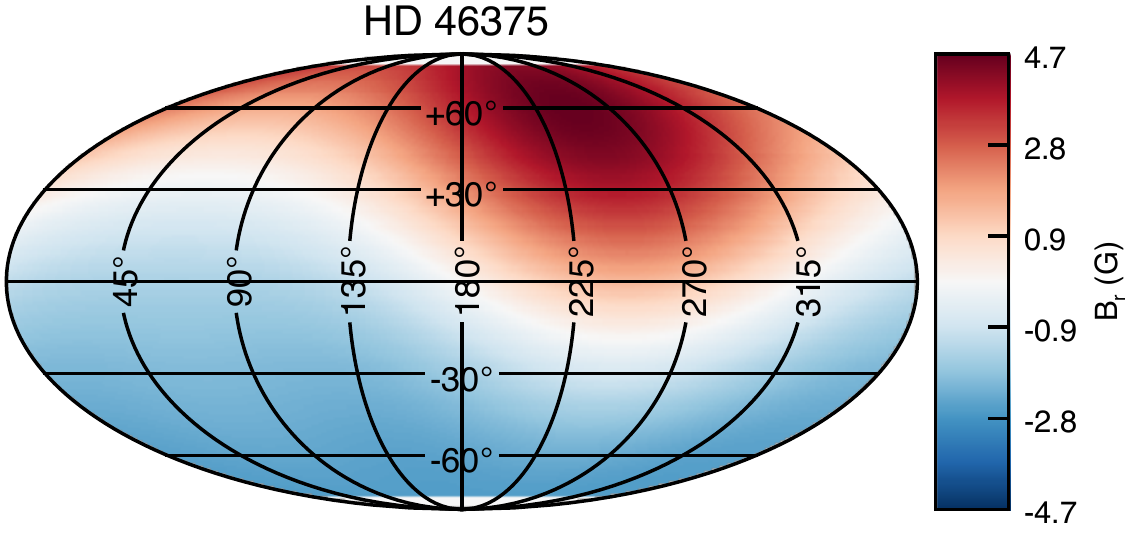}
\includegraphics[width=58mm]{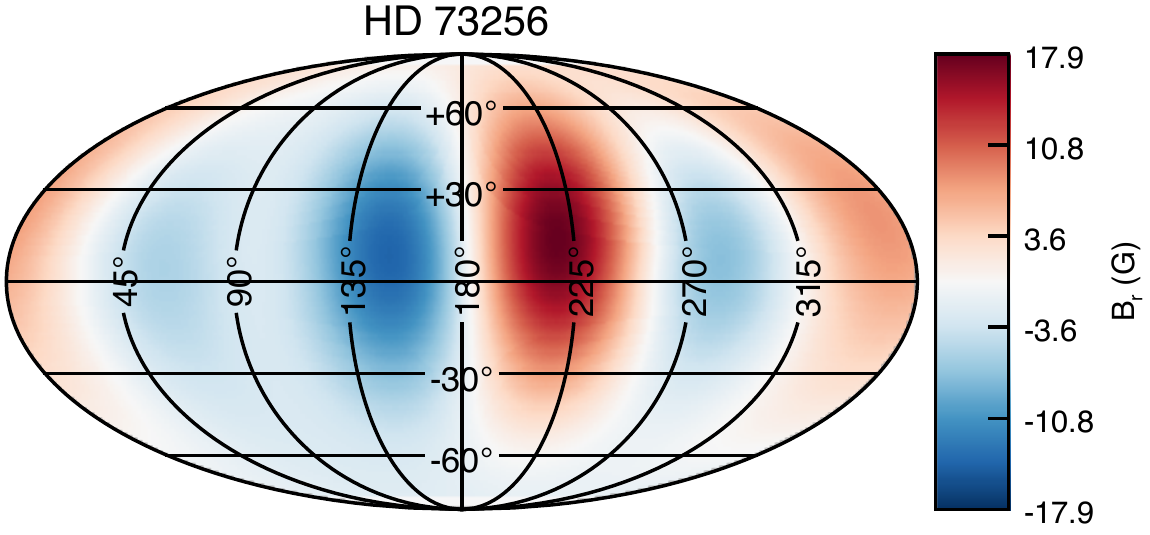}
\includegraphics[width=58mm]{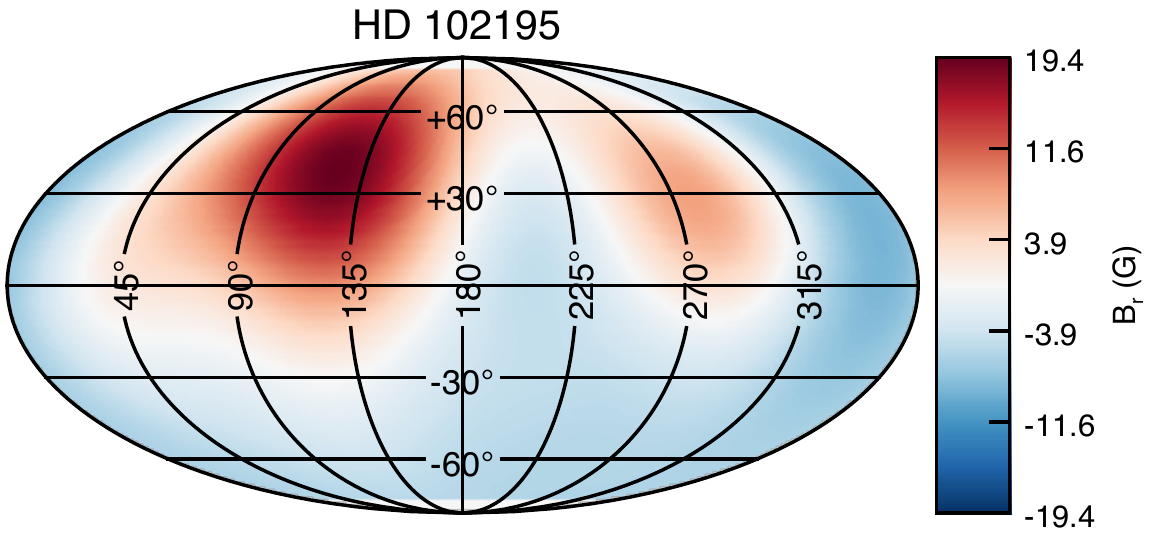}\\
\includegraphics[width=58mm]{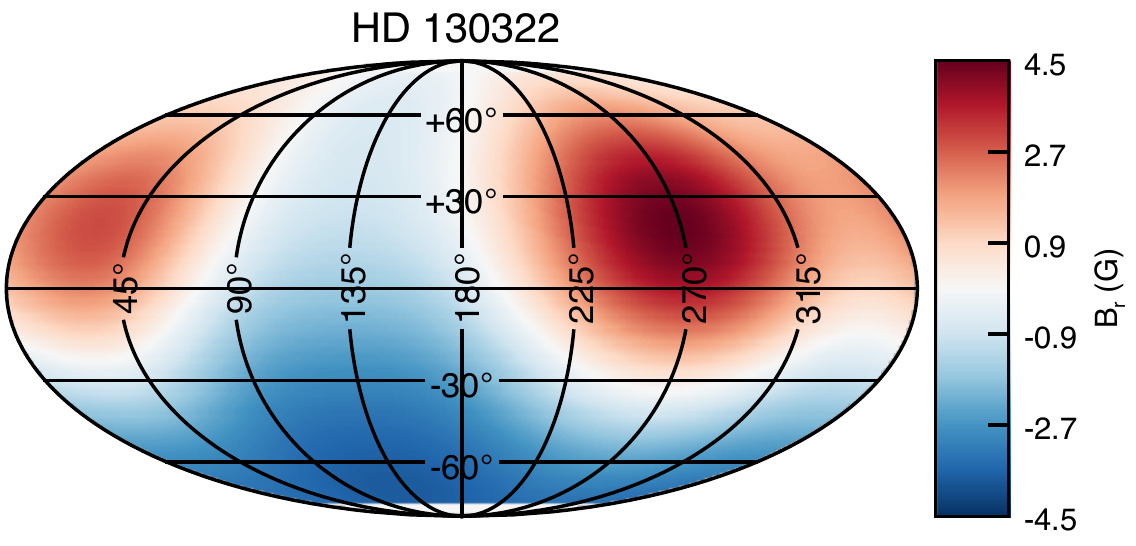}
\includegraphics[width=58mm]{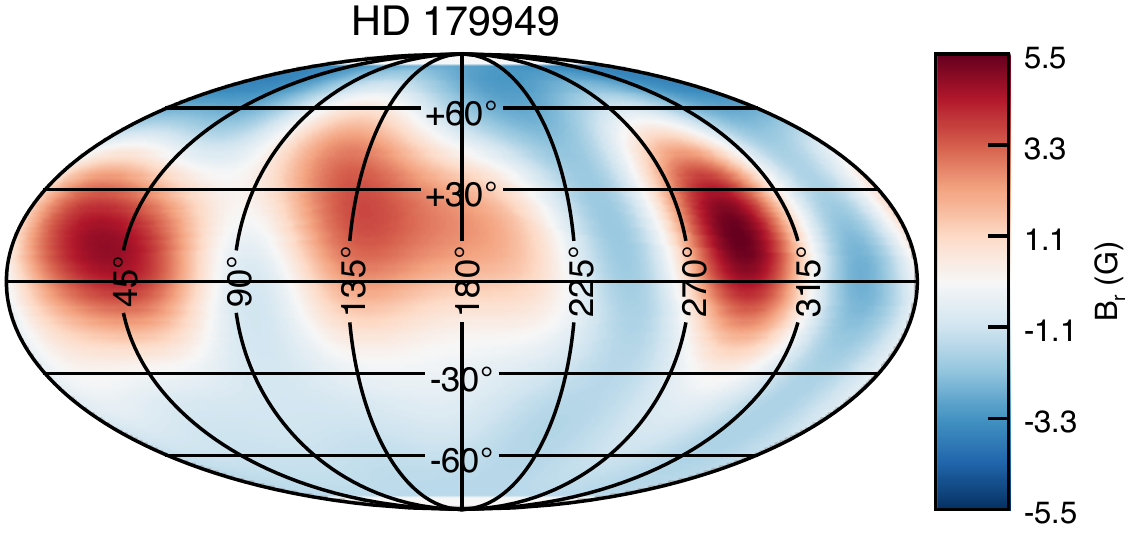}
\caption{Mollweide projection of the radial component of the stellar surface magnetic field (from \citealt{2012MNRAS.423.1006F, 2013MNRAS.435.1451F}). These observed magnetic maps are included as boundary conditions in our data-driven simulations. 
\label{fig.maps}}
\end{figure*}

The stars considered in this study consist of five solar-type stars of spectral types F8 to K1, namely: HD~46375, HD~73256, HD~102195, HD~130322 and HD~179949.  All these stars host a gaseous planet at very close orbit (i.e., a hot-Jupiter). Table \ref{tab.sample} presents a summary of the observationally-derived characteristics of the host stars and also of their hot-Jupiters (planet `b'). The large-scale surface magnetic field maps of the planet hosts have been reconstructed by \citet{2012MNRAS.423.1006F, 2013MNRAS.435.1451F} from a series of circular polarisation spectra  (acquired at CFHT/ESPaDOnS and TBL/NARVAL)  using the Zeeman-Doppler Imaging (ZDI) technique \citep[e.g.,][]{1997A&A...326.1135D,2006MNRAS.370..629D}. Figure~\ref{fig.maps} presents the radial component of the reconstructed surface field of these stars. Our targets  present surface magnetic fields with a variety of topologies and intensities. For instance, HD~46375 presents a magnetic field that is mostly dipolar, whose axis is slightly tilted with respect to the rotation axis. HD~73256, on the other hand, has a magnetic field topology that is less axisymmetric.  

\begin{table*} 
\centering
\caption{Observationally derived characteristics of the exoplanets and planet-host stars of our sample. The columns are: the host-star name, spectral type, mass ($M_\star$), radius ($R_\star$), effective temperature ($T_{\rm eff}$), rotation period ($P_{\rm rot}$), Rossby numbers (Ro), distance ($d$), inclination between the stellar rotation axis and the line-of-sight ($i$) estimated from ZDI, unsigned surface magnetic flux ($\Phi_{\rm 0}$), date of the spectropolarimetric observations, projected mass of the planet `b' ($M_p \sin i$) and semi-major axis of the planetary orbit ($a$). {For uncertainties in the quantities below, we refer the reader to the following literature.} All the values listed below were compiled by \citet{2013MNRAS.435.1451F}, except for $d$, whose references are listed in the footnote of the table, and Ro, which was derived by \citet{2014MNRAS.441.2361V} using the models of \citet{2010A&A...510A..46L}. \label{tab.sample}}    
\begin{tabular}{cccccccccccccc}  
\hline
Star	&	Spectral	&	$	M_\star	$	&	$	R_\star	$	&	$	T_{\rm eff}	$	&	$	P_{\rm rot}	$	&		Ro		&	$	d		$	&	$	i	$	&	$	\Phi_{\rm 0}	$	&		Date		&	$	M_{p} \sin i	$	&	$	a	$	\\	
ID	&	type	&	$	(M_\odot)	$	&	$	(R_\odot)	$	&		(K)		&		(d)		&				&		(pc)			&		(deg)		&		($10^{23}$ Mx)		&				&	$	(M_{\rm Jup})	$	&	$	(R_\star)	$	\\ \hline	
HD 46375	&	K1IV	&	$	0.97	$	&	$	0.86	$	&	$	5290	$	&	$	42	$	&	$	2.340	$	&	$	33.4	^1	$	&	$	45	$	&	$	0.85	$	&		2008 Jan		&	$	0.2272	$	&	$	10.0	$	\\		
HD 73256	&	G8	&	$	1.05	$	&	$	0.89	$	&	$	5636	$	&	$	14	$	&	$	0.962	$	&	$	36.5	^2	$	&	$	75	$	&	$	2.1	$	&		2008 Jan		&	$	1.869	$	&	$	9.0	$	\\		
HD 102195	&	K0V	&	$	0.87	$	&	$	0.82	$	&	$	5290	$	&	$	12.3	$	&	$	0.473	$	&	$	29.0	^3	$	&	$	50	$	&	$	2.1	$	&		2008 Jan		&	$	0.453	$	&	$	12.6	$	\\		
HD 130322	&	K0V	&	$	0.79	$	&	$	0.83	$	&	$	5330	$	&	$	26.1	$	&	$	0.782	$	&	$	30.0	^4	$	&	$	80	$	&	$	0.74	$	&		2008 Jan		&	$	1.043	$	&	$	23.2	$	\\		
HD 179949	&	F8V	&	$	1.21	$	&	$	1.19	$	&	$	6168	$	&	$	7.6	$	&	$	>1.726	$	&	$	27.0	^5	$	&	$	60	$	&	$	1.3	$	&		2007 Jun		&	$	0.902	$	&	$	7.9	$	\\ \hline		
\end{tabular}\\
$^1$\citet{2000ApJ...536L..43M}; $^2$\citet{2003A&A...407..679U}; $^3$\citet{2006ApJ...648..683G}; $^4$\citet{2000A&A...356..590U}; $^5$\citet{2007A&A...475..519H}
\end{table*}

\section{Stellar wind model}\label{sec.model}
The stellar wind model we use here is identical to the one presented in \citet{2014MNRAS.438.1162V}. We use the 3D MHD numerical code BATS-R-US \citep{1999JCoPh.154..284P,2012JCoPh.231..870T} to simulate the stellar winds. BATS-R-US solves the set of ideal MHD equations for the mass density $\rho$, the plasma velocity ${\bf u}=\{ u_r, u_\theta, u_\varphi\}$, the magnetic field ${\bf B}=\{ B_r, B_\theta, B_\varphi\}$, and the gas pressure $P$:
\begin{equation}
\label{eq:continuity_conserve}
\frac{\partial \rho}{\partial t} + \boldsymbol\nabla\cdot \left(\rho {\bf u}\right) = 0,
\end{equation}
\begin{equation}
\label{eq:momentum_conserve}
\frac{\partial \left(\rho {\bf u}\right)}{\partial t} + \boldsymbol\nabla\cdot\left[ \rho{\bf u\,u}+ \left(P + \frac{B^2}{8\pi}\right)I - \frac{{\bf B\,B}}{4\pi}\right] = \rho {\bf g},
\end{equation}
\begin{equation}
\label{eq:bfield_conserve}
\frac{\partial {\bf B}}{\partial t} + \boldsymbol\nabla\cdot\left({\bf u\,B} - {\bf B\,u}\right) = 0,
\end{equation}
\begin{equation}
\label{eq:energy_conserve}
\frac{\partial\varepsilon}{\partial t} +  \boldsymbol\nabla \cdot \left[ {\bf u} \left( \varepsilon + P + \frac{B^2}{8\pi} \right) - \frac{\left({\bf u}\cdot{\bf B}\right) {\bf B}}{4\pi}\right] = \rho {\bf g}\cdot {\bf u} ,
\end{equation}
where
\begin{equation}\label{eq:energy_density}
\varepsilon=\frac{\rho u^2}{2}+\frac{P}{\gamma-1}+\frac{B^2}{8\pi} .
\end{equation}
We assume the wind is polytropic, in which $P\propto \rho^\gamma$ and $\gamma$ is the polytropic index. To derive the temperature, we consider an ideal gas, so $P=n k_B T$, where  $k_B$ is the Boltzmann constant, $T$ is the temperature, $n=\rho/(\mu m_p)$ is the particle number density of the stellar wind, $\mu m_p$ is the mean mass of the particle. In this work, we adopt $\gamma=1.1$, similar to the effective adiabatic index measured in the solar wind \citep{2011ApJ...727L..32V}, and $\mu=0.5$, for a fully ionised hydrogen plasma.

At the initial state of the simulations, we assume that the wind is thermally driven \citep{1958ApJ...128..664P}. The stellar rotation period $P_{\rm rot}$, $M_\star$ and $R_\star$ are given in Table \ref{tab.sample}. At the base of the corona ($r=R_\star$), we adopt a wind coronal temperature $T_0 = 2\times 10^6$~K and wind number density $n_0=10^{9}$cm$^{-3}$ (Section~\ref{sec.limitations} discusses the choices of $n_0$ and $T_0$ and how they affect our results). With this numerical setting, the initial solution for the density, pressure (or temperature) and wind velocity profiles are fully specified.  The radial component of the magnetic field $B_r$ anchored at the base of the wind, is reconstructed from observations (Fig.~\ref{fig.maps}). The other two components of the surface field are assumed to be potential ($\boldsymbol\nabla \times {\bf B}=0$), as it has been shown that stellar winds are largely unaffected by the non-potential part of the observed surface field \citep{2013MNRAS.431..528J}. At the initial state, we assume that the field considered in the simulation box is potential up to a radial distance $r=r_{\rm SS}$ (known as the source surface) and, beyond that,  the magnetic field lines are considered to be open and purely radial. As the simulation evolves in time, the wind particles interact with the magnetic field lines (and vice-versa), removing the field from its initial potential state. For all the cases studied here, we take $r_{\rm SS}=4~R_\star$, but we note that different values of $r_{\rm SS}$ produce similar final steady-state solutions for the simulations \citep{2011MNRAS.412..351V, 2014MNRAS.438.1162V}. 

Once set at the initial state of the simulation, the values of the observed $B_r$ are held fixed at the base of the wind throughout the simulation run, as are the coronal base density and thermal pressure. A zero radial gradient is set to the remaining components of ${\bf B}$ and ${\bf u}=0$ in the frame corotating with the star. The outer boundaries at the edges of the grid have outflow conditions. 
The rotation axis of the star is aligned with the $z$-axis, and the star is assumed to rotate as a solid body. Our grid is Cartesian and the star is placed at the origin of the grid, which extends in $x$, $y$, and $z$ from $-20$ to $20~R_\star$, except for HD~102195, whose simulation box extends from $-24$ to $24~R_\star$, as to extend out to the orbit of the planet. BATS-R-US uses block adaptive mesh refinement. The finest resolved cells are located close to the star (for $r \lesssim 2~R_\star$), where the linear size of the cubic cell is $0.0097~R_\star$ (or $0.012~R_\star$ for the simulation of HD~102195). The coarsest cell  has a linear size of $0.31~R_\star$ (or $0.37~R_\star$ for HD~102195) and is located at the outer edges of the grid. The total number of cells in our simulations is around $40$ million. As the simulations evolve in time, both the wind and magnetic field lines are allowed to interact with each other. The resultant solution, obtained self-consistently, is found when the system reaches steady state in the reference frame corotating with the star.  

\section{Derived properties of the stellar winds}\label{sec.results}

Table~\ref{tab.results} presents the properties of the stellar winds obtained in our simulations. The unsigned observed surface magnetic flux is 
\begin{equation}\label{eq.phi0}
\Phi_0 = \oint_{S_\star} |B_r (R_\star, \theta, \varphi)| {\rm d} S_\star
\end{equation}
and the unsigned open magnetic flux is
\begin{equation} \label{eq.phiopen}
\Phi_{\rm open} = \oint_{S_{\rm sph}} |B_r (r, \theta, \varphi)| {\rm d} S_{\rm sph}. 
\end{equation}
The surface flux (Table~\ref{tab.sample}) is integrated over the surface of the star $S_\star$ and the open flux  (Table~\ref{tab.results}) over a spherical surface $S_{\rm sph}$ at a distance $r$ from the star, where all the magnetic field lines are open. The mass-loss rate $\dot{M}$ of the stellar wind, which outflows along open magnetic field lines, can be calculated as the flux of mass integrated across $S_{\rm sph}$ 
\begin{equation}\label{eq.mdot}
\dot{M} = \oint \rho u_r {\rm d} S_{\rm sph},
\end{equation}
where $\dot{M}$ is a constant of the wind. Similarly, the angular momentum loss rates can be calculated as the angular momentum flux across $S_{\rm sph}$ 
\begin{equation}\label{eq.jdot}
\dot{J} = \oint_{S_{\rm sph}} \left[  - \frac{\varpi B_\varphi B_r}{4 \pi} + \varpi u_\varphi \rho u_r \right]  {\rm d} S _{\rm sph} 
 \end{equation} 
\citep{1970MNRAS.149..197M,1999stma.book.....M,2014MNRAS.438.1162V}, where $\varpi=(x^2+y^2)^{1/2}$ is the cylindrical radius. In our simulations, we find that $\mdot$ ranges from $\sim 2$ to $8 \times 10^{-13} ~\msano$ and $\jdot$ between $\sim 0.14$ and $2.4 \times 10^{31}$~erg for the stars in our sample. The open flux ranges from $26\%$ to $69\%$ of the large-scale unsigned surface flux. These values are within the range $(4.4 - 8.4) \times 10^{22}$~Mx. For the solar wind,  \citet{2006ApJ...644..638W} obtained magnetic field values at the orbit of the Earth in the range between $0.01$ and $0.05$~mG, or in terms of open magnetic fluxes, in the range $(2.8 - 14) \times 10^{22}$~Mx, depending on the phase of the solar activity cycle. Although the range of open fluxes calculated for the simulations presented here fall within the values of the solar wind, we show in Section~\ref{sec.planets} that the values of the interplanetary magnetic field at the orbits of the hot-Jupiters are more than $100$ times larger than the interplanetary magnetic field at the Earth's orbit (compare $0.01$ to $0.05$~mG to the values presented in Table~\ref{tab.resultsp}). 

\begin{table*}
\centering
\caption{Characteristics of the stellar winds. The columns are: the star name, the stellar wind mass-loss rate ($\mdot$), angular momentum-loss rate ($\dot{J}$),  unsigned open magnetic flux ($\Phi_{\rm open}$), average radii of the \alf\ surfaces ($\langle r_A \rangle$), its minimum and maximum value ($r_A^{\rm min}$ and $r_A^{\rm max}$) and the effective radius of the source surface derived from the MHD models ($r_{\rm ss}^{\rm eff}$). In our simulations, $\mdot$, $\jdot$ and $\Phi_{\rm open}$ are conserved within $0.03\%$, $3\%$ and $4\%$, respectively. \label{tab.results}}    
\begin{tabular}{cccccccccccccc}  
\hline
Star	&	$	\dot{M}	$	&	$	\dot{J}	$	&	$	\Phi_{\rm open}	$	&	$			\langle r_A \rangle			$	&	$	[	r_A^{\rm min}	,	r_A^{\rm max}	]	$&	$	r_{\rm SS}^{\rm eff}	$	\\
ID	&	$	(10^{-13} M_\odot~\rm{yr}^{-1})	$	&		($10^{31}$ erg)		&	$	(\Phi_{\rm 0})	$	&	$			(R_\star)			$	&	$			(R_\star)			$	&	$	(R_\star)	$	\\ \hline
HD 46375	&	$	1.9	$	&	$	0.14	$	&	$	0.52	$	&	$			5.1			$	&	$	[	3.0	,	6.1	]	$	&	$	2.7	$	\\
HD 73256	&	$	2.1	$	&	$	2.3	$	&	$	0.26	$	&	$			6.2			$	&	$	[	1.8	,	8.1	]	$	&	$	5.6	$	\\
HD 102195	&	$	3.2	$	&	$	2.0	$	&	$	0.41	$	&	$			6.4			$	&	$	[	2.3	,	7.5	]	$	&	$	5.6	$	\\
HD 130322	&	$	5.8	$	&	$	0.36	$	&	$	0.69	$	&	$			3.5			$	&	$	[	1.6	,	4.2	]	$	&	$	1.9	$	\\
HD 179949	&	$	8.0	$	&	$	2.4	$	&	$	0.34	$	&	$			2.8			$	&	$	[	1.0	,	3.7	]	$	&	$	3.0	$	\\ \hline	
\end{tabular}
\end{table*}

The left panels in Figure~\ref{fig.IC-SS} show the final configuration of the magnetic field lines obtained through self-consistent interaction between magnetic and wind forces after the simulations reached steady state. Although we assume the magnetic field is current-free in the initial state of our simulations, this configuration is deformed when the interaction of the wind particles with the magnetic field lines (and vice-versa) takes place (currents are created in the system). The right panels of Figure~\ref{fig.IC-SS}  show the Alfv\'en surface $S_A$ of each simulation. This surface is defined as the location where the wind velocity reaches the local Alfv\'en velocity ($v_A = B(4 \pi \rho)^{-1/2}$). Inside $S_A$, where the magnetic forces dominate over the wind inertia, the stellar wind particles are forced to follow the magnetic field lines. Beyond $S_A$, the wind inertia dominates over the magnetic forces and, as a consequence, the magnetic field lines are dragged by the stellar wind. In models of stellar winds, the \alf\ surface has an important property for the characterisation of angular momentum losses, as it defines the lever arm of the torque that the wind exerts on the star \citep[e.g.,][]{1967ApJ...148..217W}. Its location is also relevant in the studies of magnetic interactions with planets \citep[e.g.,][]{2014ApJ...795...86S,2014ApJ...790...57C}. As shown in \citet{2014MNRAS.438.1162V}, the \alf\ surfaces of the objects investigated here have irregular, asymmetric shapes as a consequence of the irregular distribution of the observed magnetic field. To illustrate the difference in sizes of these surfaces, we show in the right panels of Figure~\ref{fig.IC-SS} the scales of the images plotted (red lines). We find that the average radius of the \alf\ surfaces range between $2.8~R_\star$ (for HD~179949) to $6.4~R_\star$ (for HD~102195).

\begin{figure*}
\includegraphics[width=70mm]{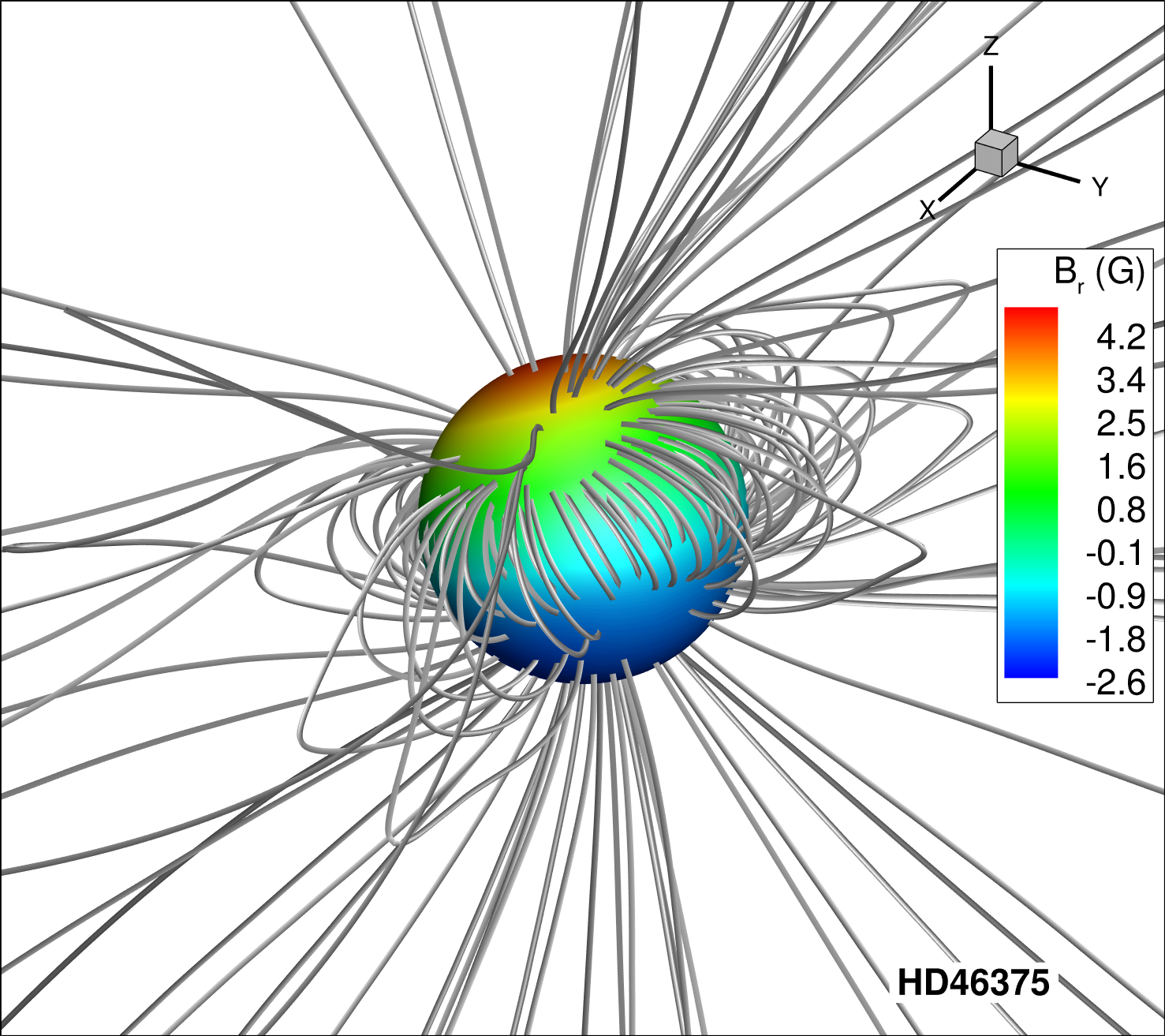}
\includegraphics[width=70mm]{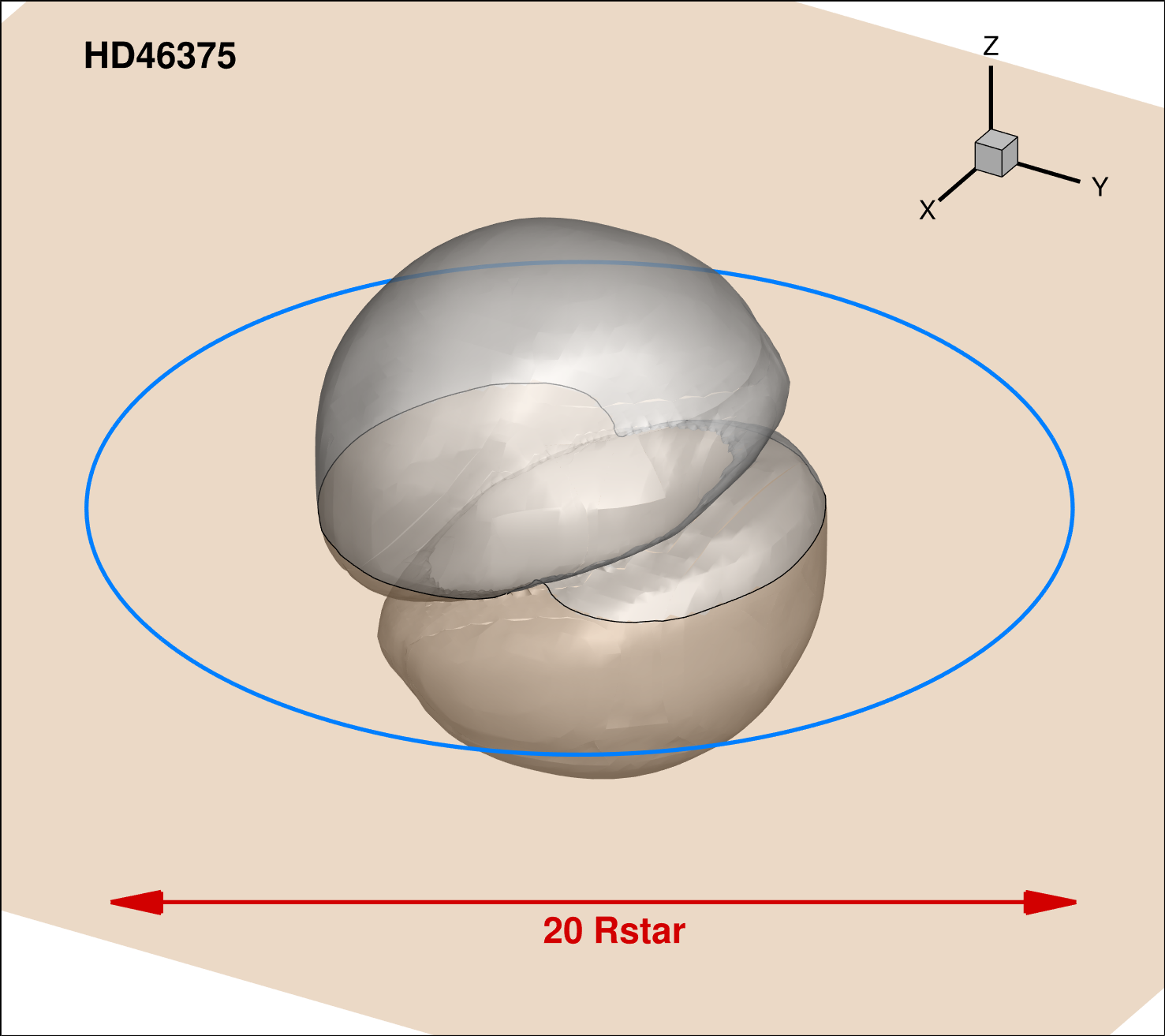}\\
\includegraphics[width=70mm]{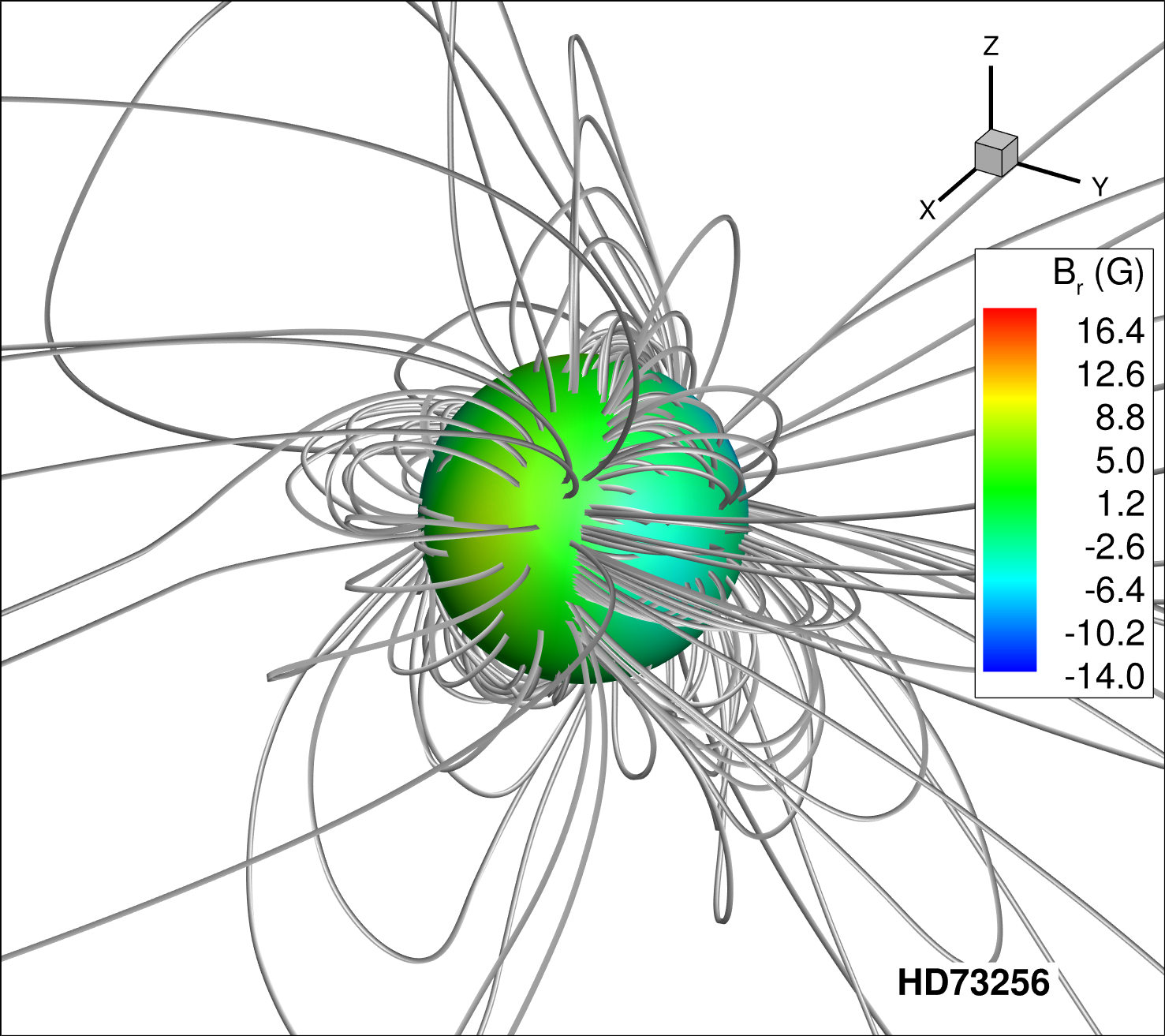}
\includegraphics[width=70mm]{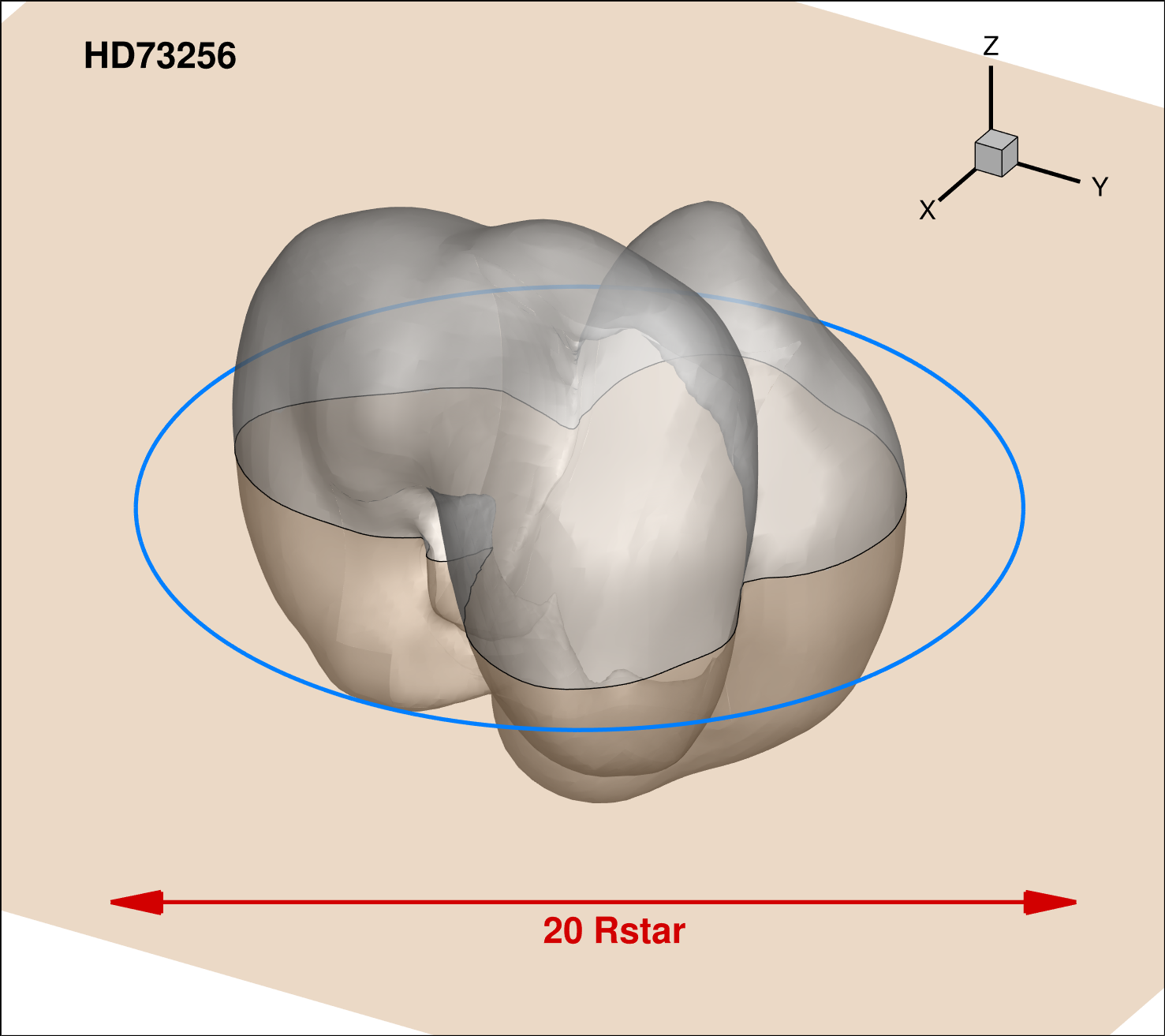}\\
\includegraphics[width=70mm]{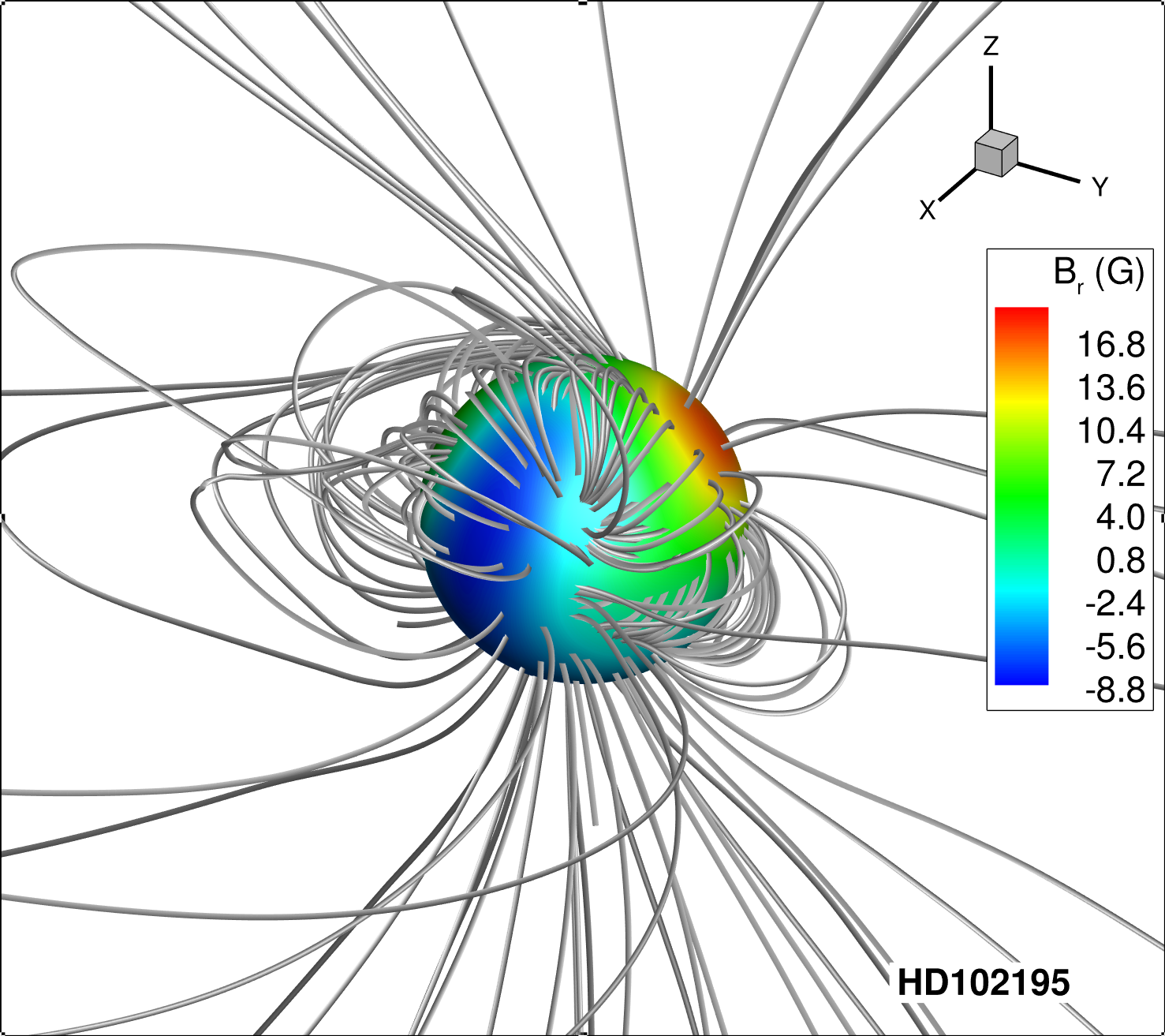}
\includegraphics[width=70mm]{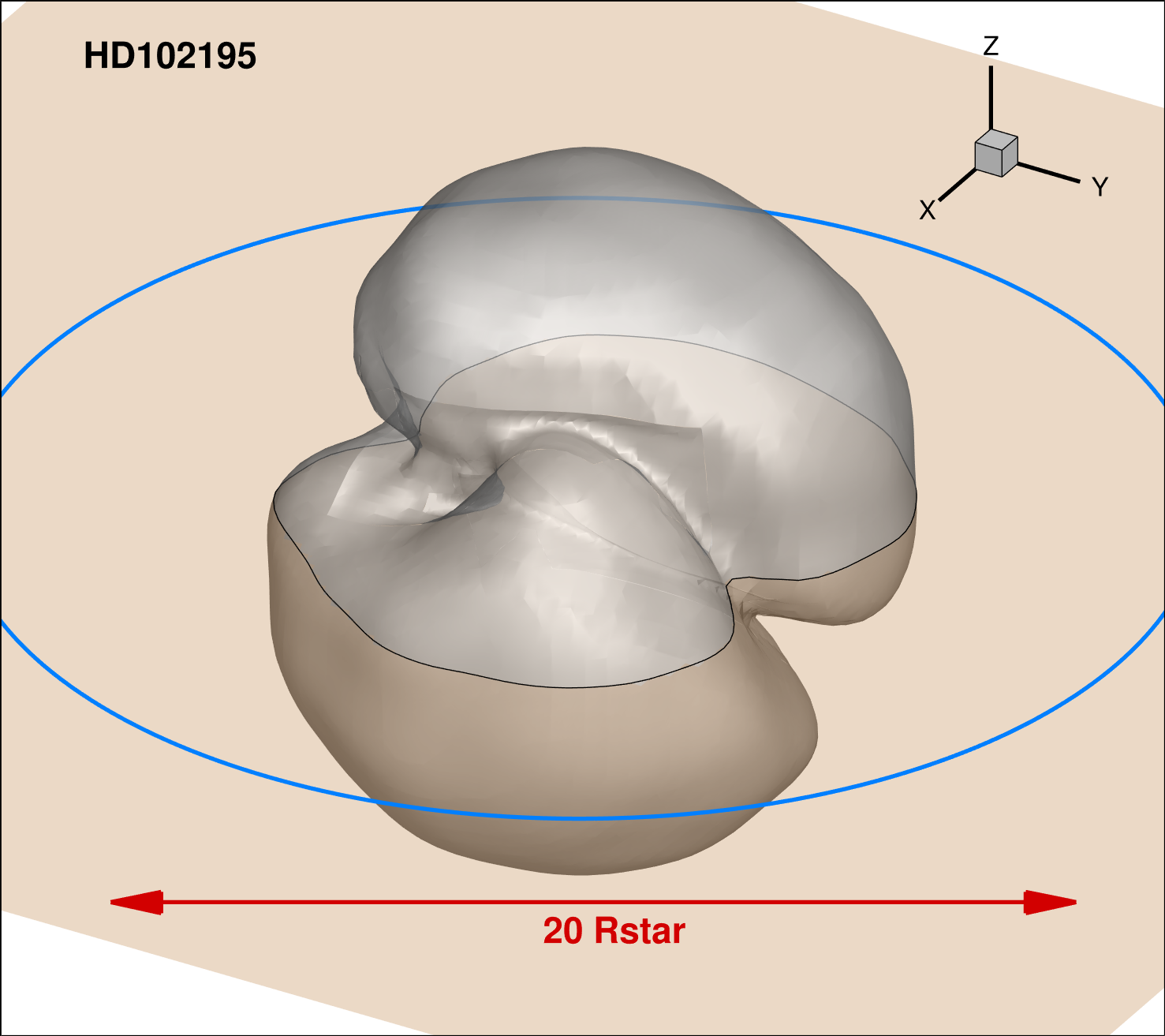}
\caption{Left: The final configuration of the magnetic field lines after the wind solution has relaxed in the grid. Over-plotted at the surface of the star is the observationally reconstructed stellar magnetic field \citep{2012MNRAS.423.1006F,2013MNRAS.435.1451F}, used as boundary condition for the radial magnetic field.  Right: The \alf\ surfaces are shown in grey. Note their irregular, asymmetric shapes due to the irregular distribution of the observed field. The equatorial ($xy$) planes of the star, assumed to contain the orbits of the planet, are also shown, as are the intersections between the $xy$ plane and the \alf\ surface (thin black contour) and the orbital radius of the planet (thick blue contour).  \label{fig.IC-SS}}
\end{figure*}

\begin{figure*}
\includegraphics[width=70mm]{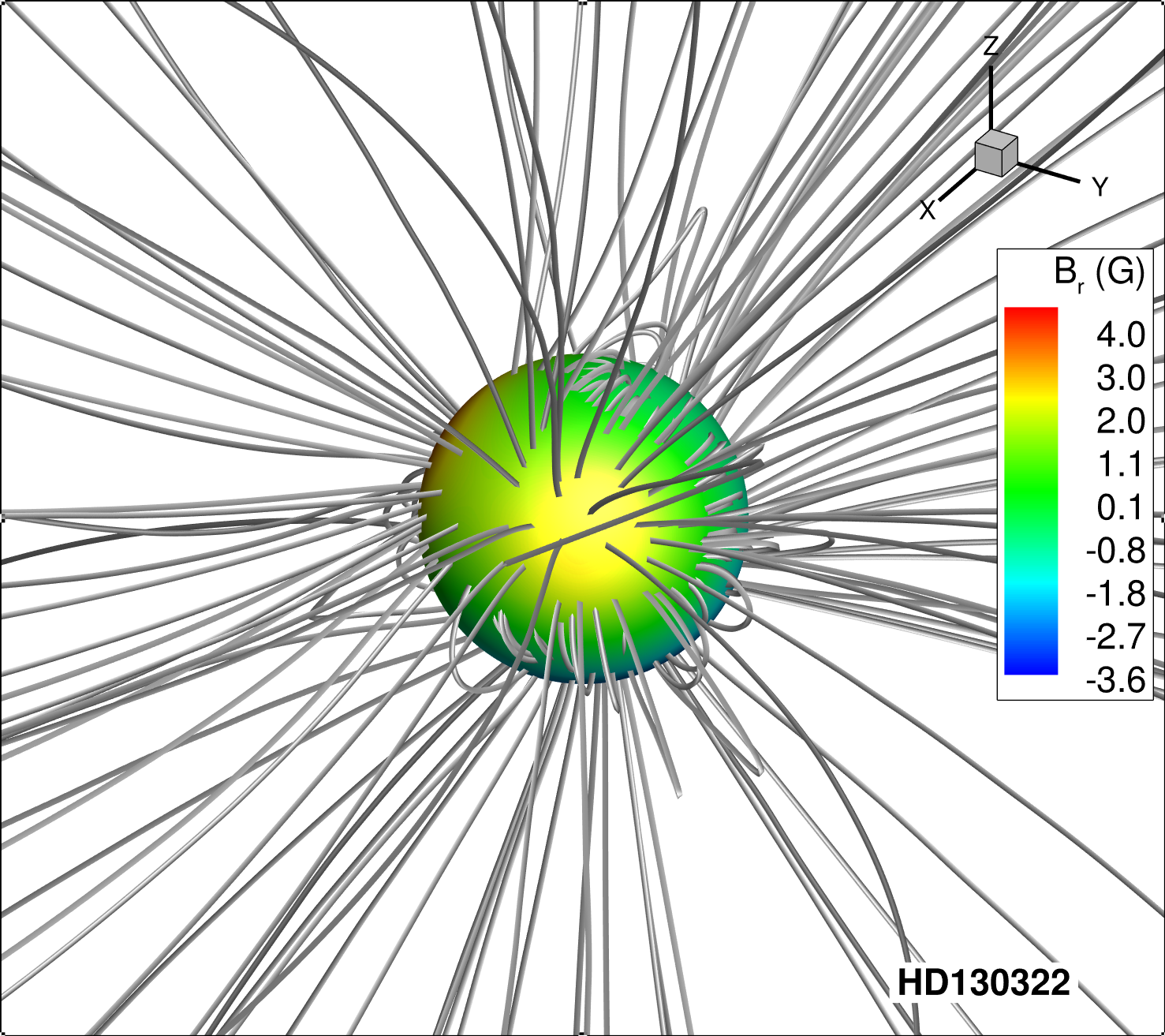}
\includegraphics[width=70mm]{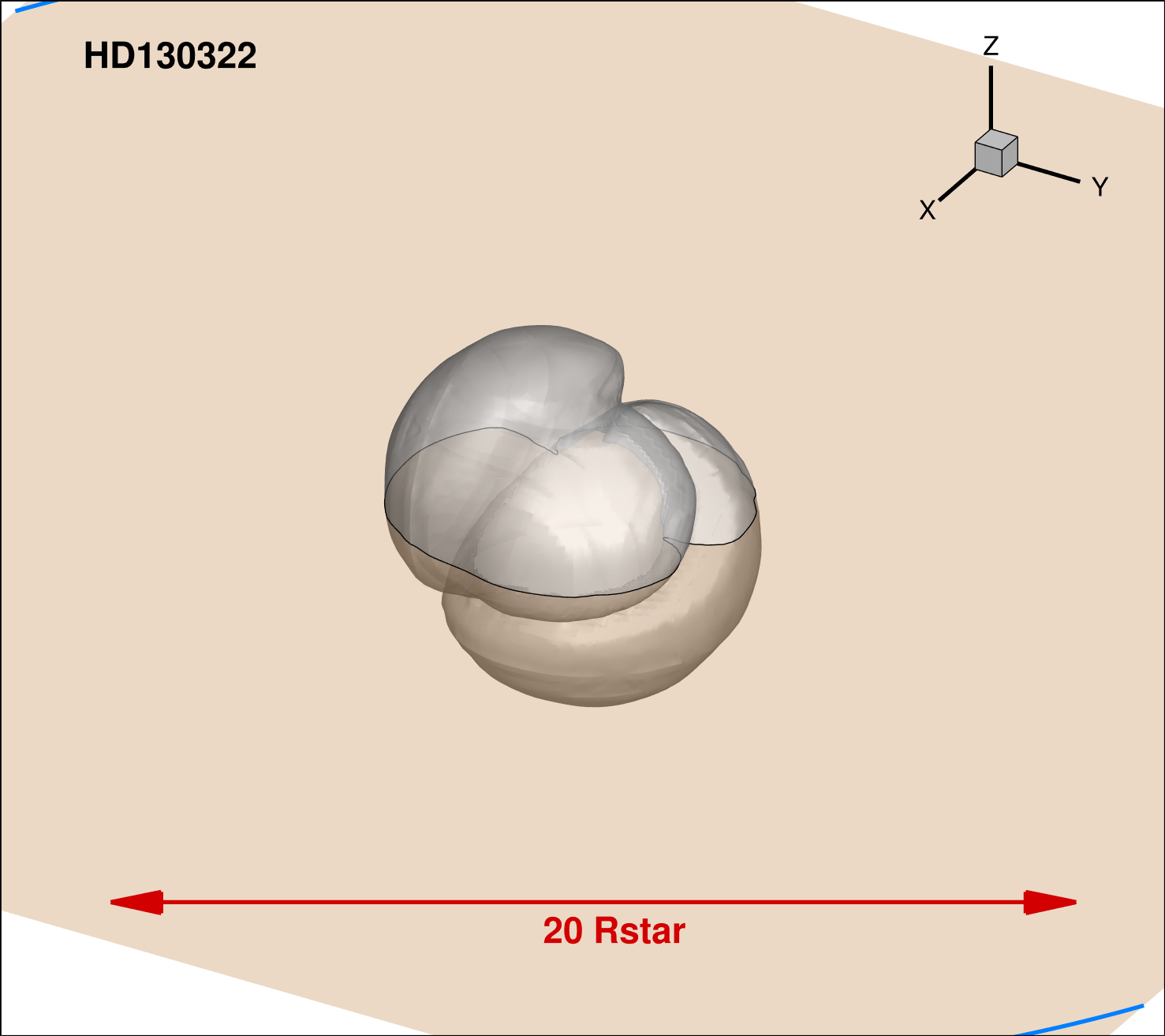}\\
\includegraphics[width=70mm]{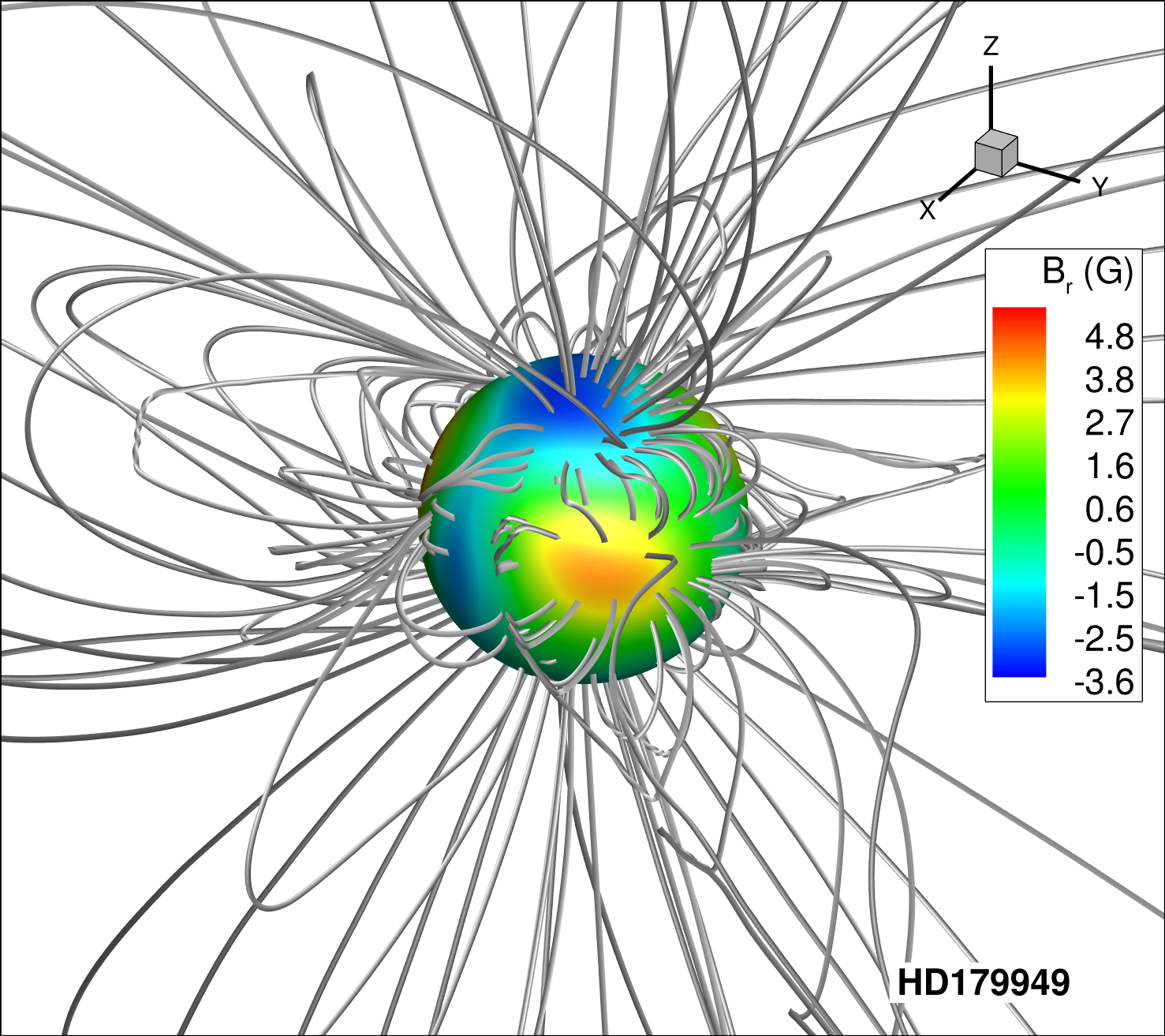}
\includegraphics[width=70mm]{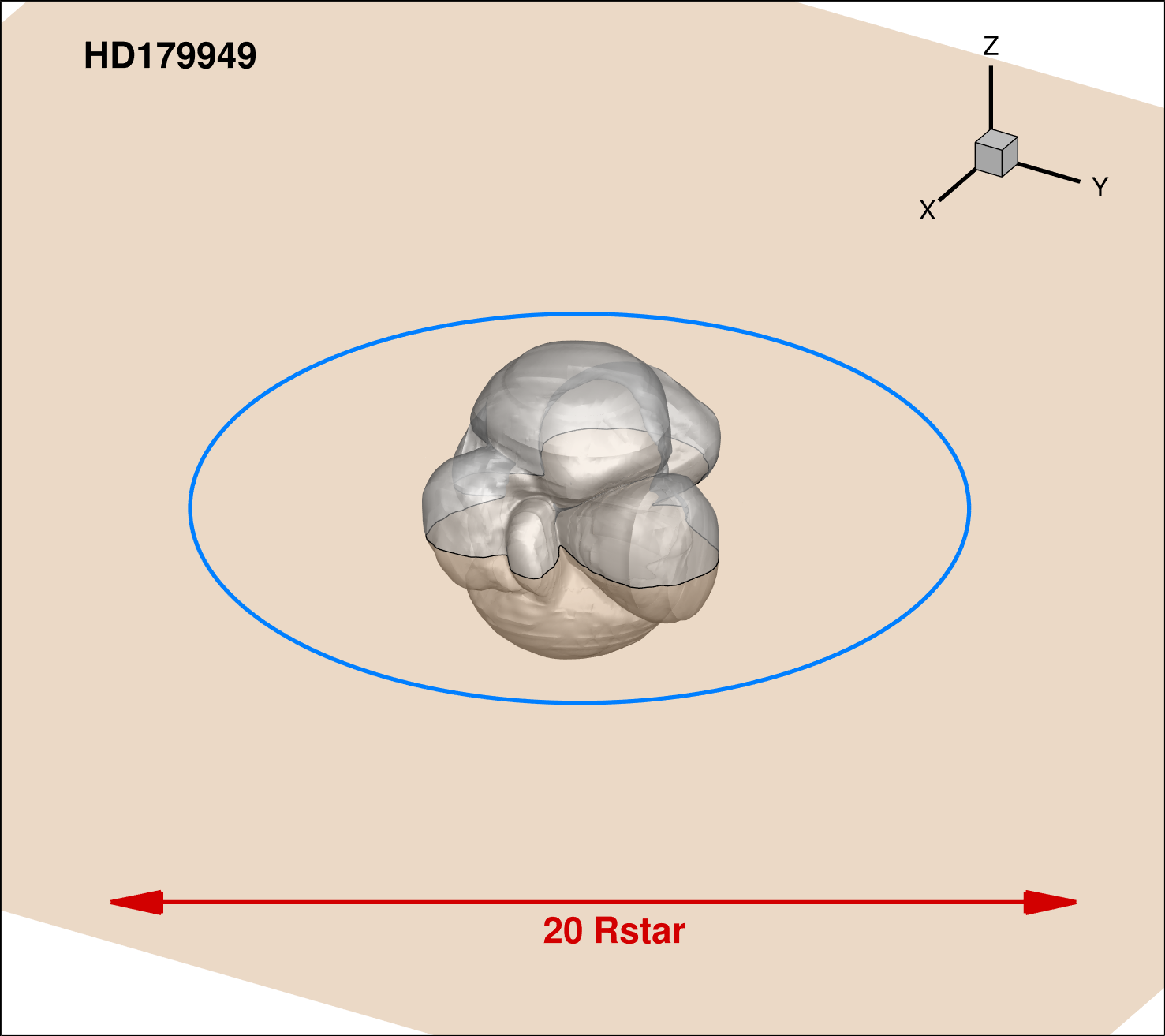}\\
\contcaption{}
\end{figure*}

In order to provide constraints for analytical methods of extrapolation of magnetic field lines, we also compute here the MHD equivalent of the source surface. In particular, the potential field source surface (PFSS) method has proven to be a fast and simple way to extrapolate surface magnetic fields into the stellar coronal region \citep{1999MNRAS.305L..35J,2002MNRAS.333..339J, 2013A&A...557A..67V}. It is also used here as the initial conditions for our simulations. However, the PFSS method has an unconstrained parameter: the radius $r_{\rm SS}$ of the source surface, beyond which the magnetic field lines are assumed open and purely radial, as a way to mimic the effects of a stellar wind. Because of stellar rotation and magnetic field stresses, in the MHD solutions, the surface where all magnetic field lines are purely radial does not exist -- even in the region of open field lines, there is always $B_\theta$ and, especially, $B_\varphi$ components that are non-null. Therefore, we define here an ``effective radius of the source surface'' $r_{\rm SS}^{\rm eff}$ as  the radius of the spherical surface where $97$~percent of the average magnetic field is contained in the radial component (i.e., $\langle |B_r| \rangle / \langle |B| \rangle =0.97$, based on \citealt{2006ApJ...653.1510R}). For some of the stars in our sample (HD~73256 and HD~179949), the ratio $\langle |B_r| \rangle / \langle |B| \rangle$ does not reach the 97-percent level and in such cases, we take $r_{\rm SS}^{\rm eff}$ to be the position where $\langle |B_r| \rangle / \langle |B| \rangle$ is maximum. Table~\ref{tab.results} shows that $r_{\rm SS}^{\rm eff}$ is in the range between $1.9~R_\star$ and $5.6~R_\star$, indicating a compact region of closed field lines. We note that this size is similar to the usual adopted size of $2.5~R_\odot$ from PFSS methods of the solar coronal magnetic field and also similar to the values obtained in other MHD simulations of winds \citep{2006ApJ...653.1510R,2011MNRAS.412..351V,2014MNRAS.438.1162V}.

\section{Characterising the local environment surrounding hot-Jupiters and resultant interactions}\label{sec.planets}
All the stars in our sample host giant planets orbiting at close distances. Mercury, the closest planet to our Sun, has a semimajor orbital axis of about 0.39~au, or equivalently, of about $83~R_\odot$. The hot-Jupiters in our sample have considerably closer orbits, with semimajor axes of about $9$ to $23~R_\star$ (i.e., about $9$ to $4$ times closer than Mercury). As a consequence, the hot-Jupiters in our sample interact with much denser winds that have larger ram pressures than those typically found around the planets in the solar System. In addition, because the hot-Jupiters are located much closer to the star, the large-scale magnetic field at the orbit of these planets has also a larger strength compared to the interplanetary magnetic field strength of our solar System planets.

\begin{figure*}
\includegraphics[height=53mm]{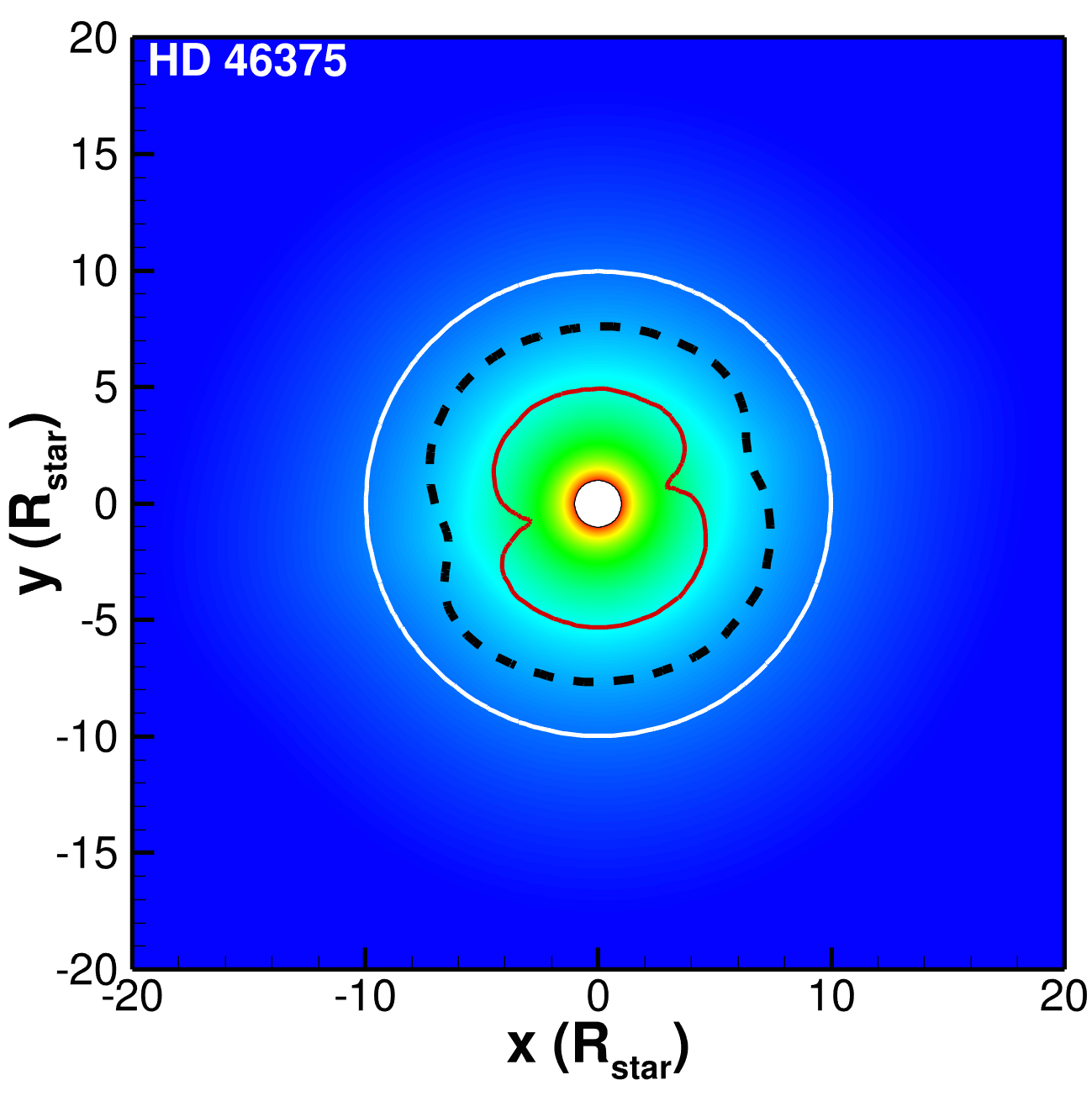}
\includegraphics[height=53mm]{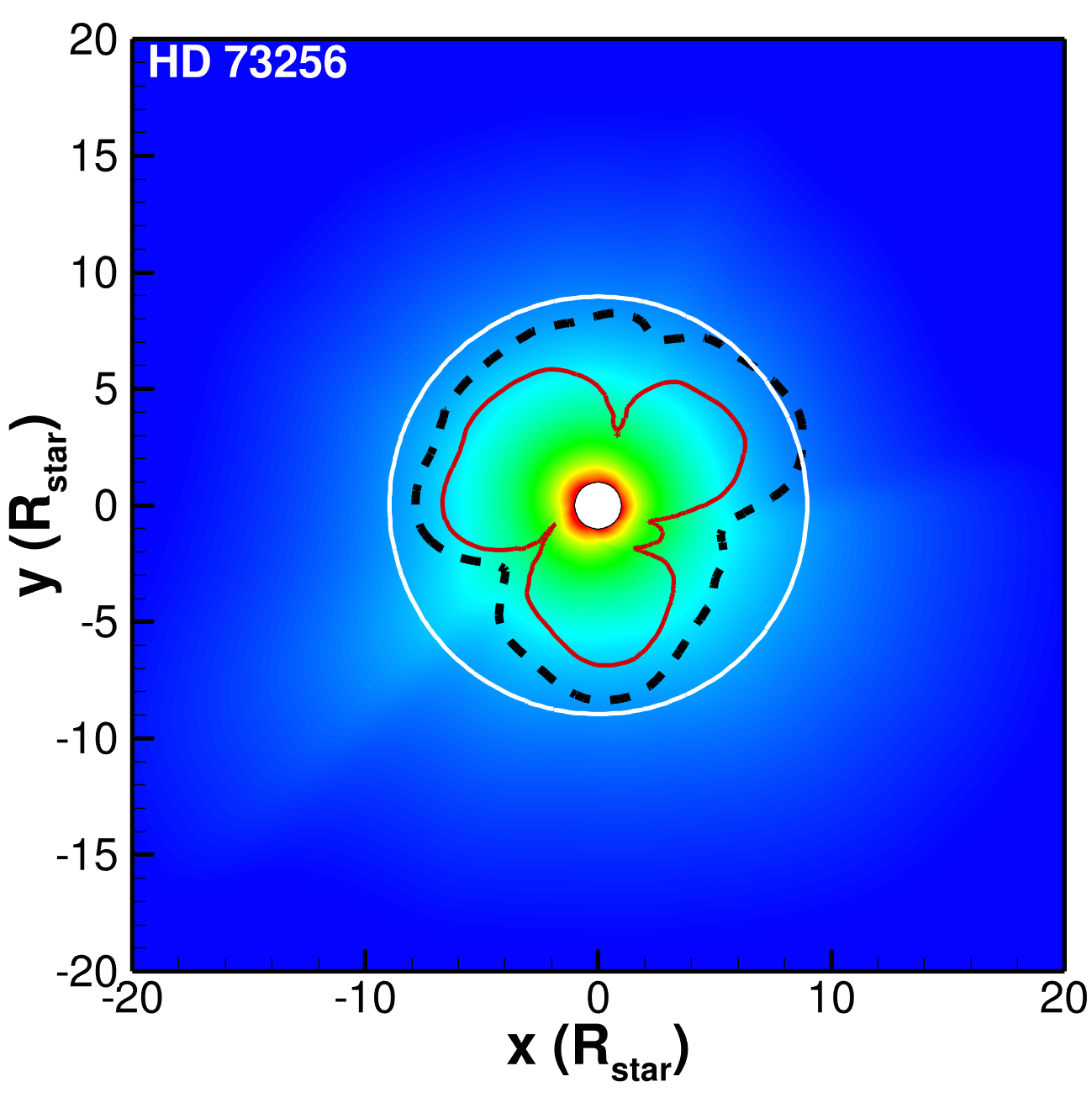}
\includegraphics[height=53mm]{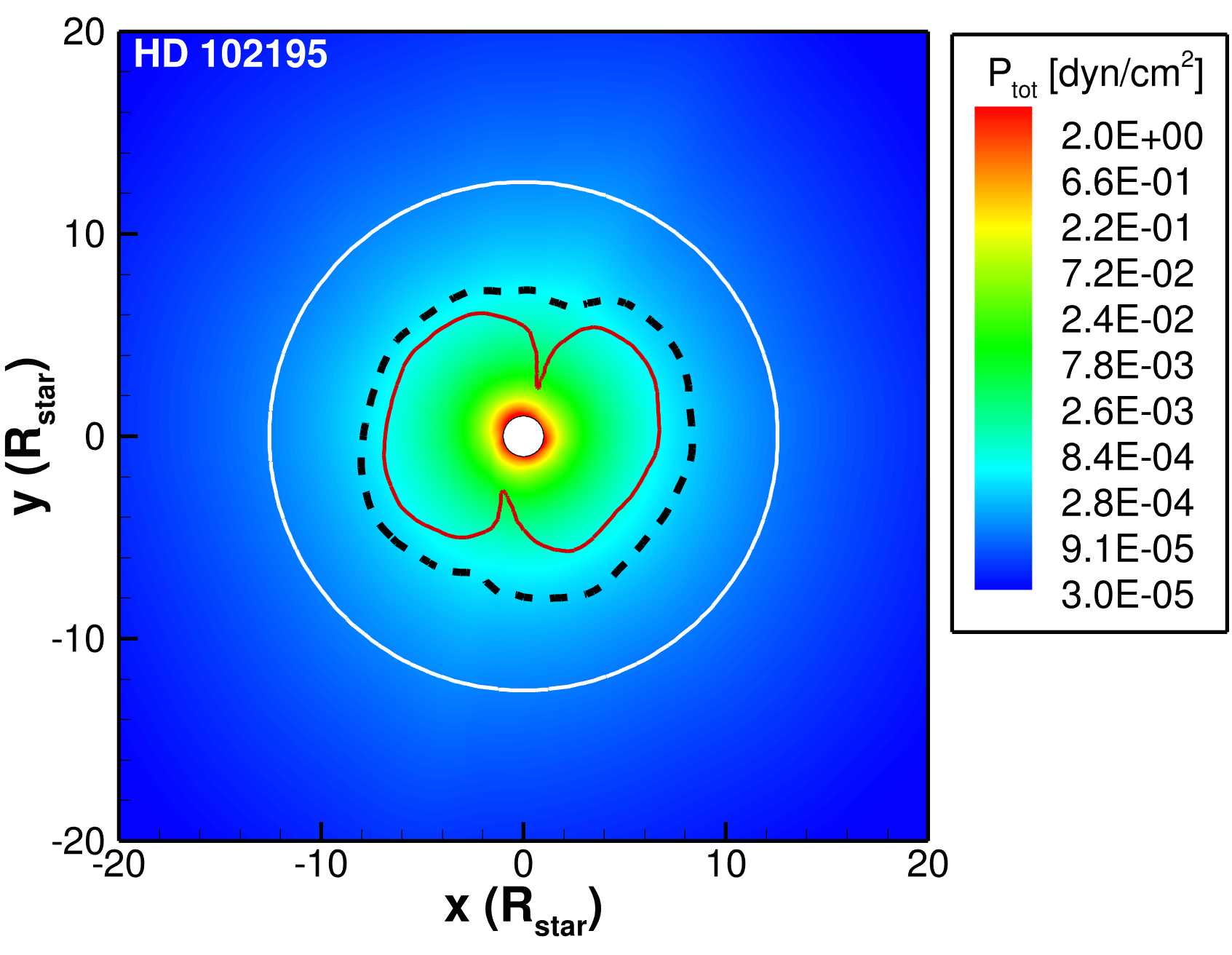}
\\
\includegraphics[height=53mm]{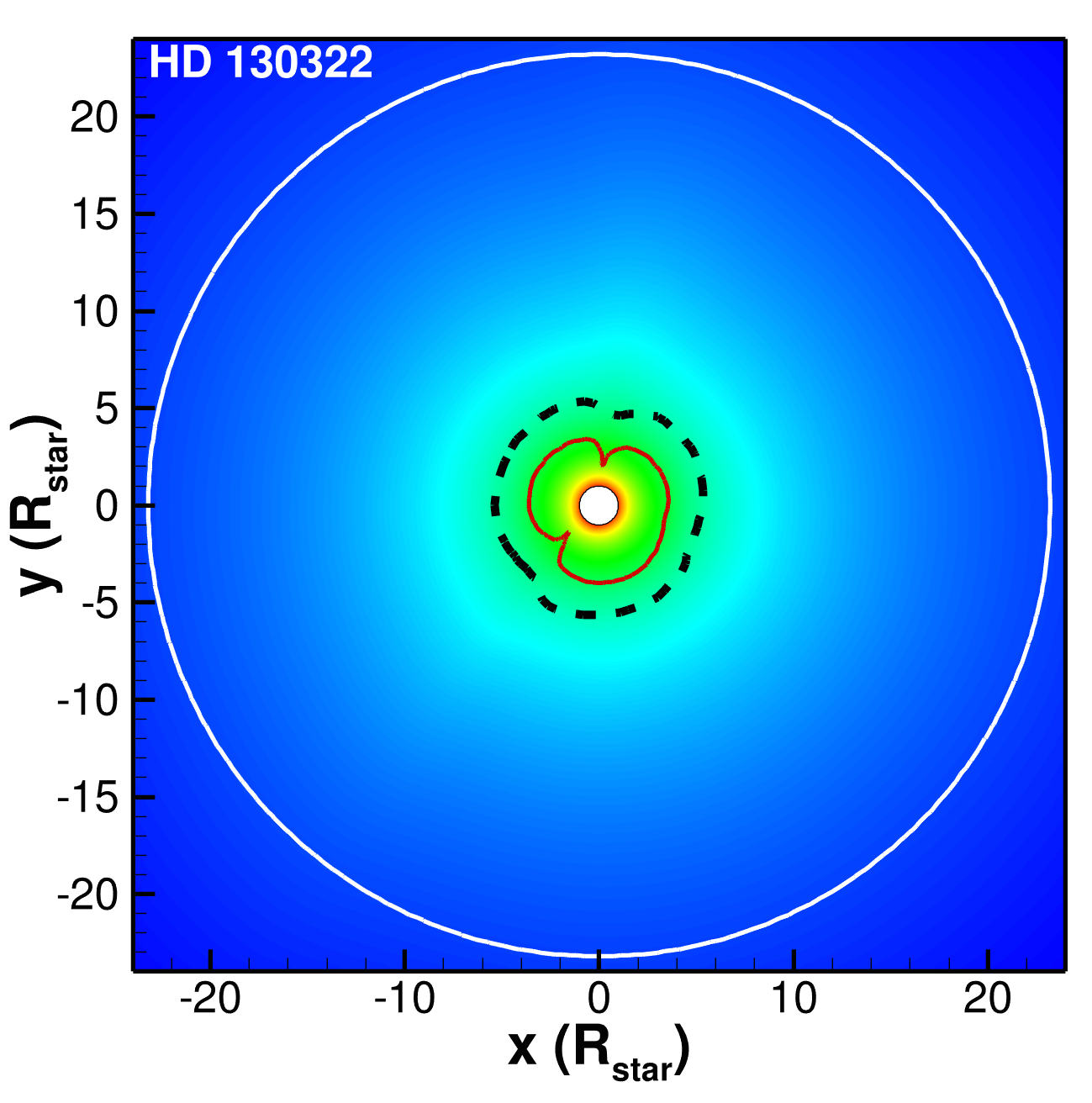}
\includegraphics[height=53mm]{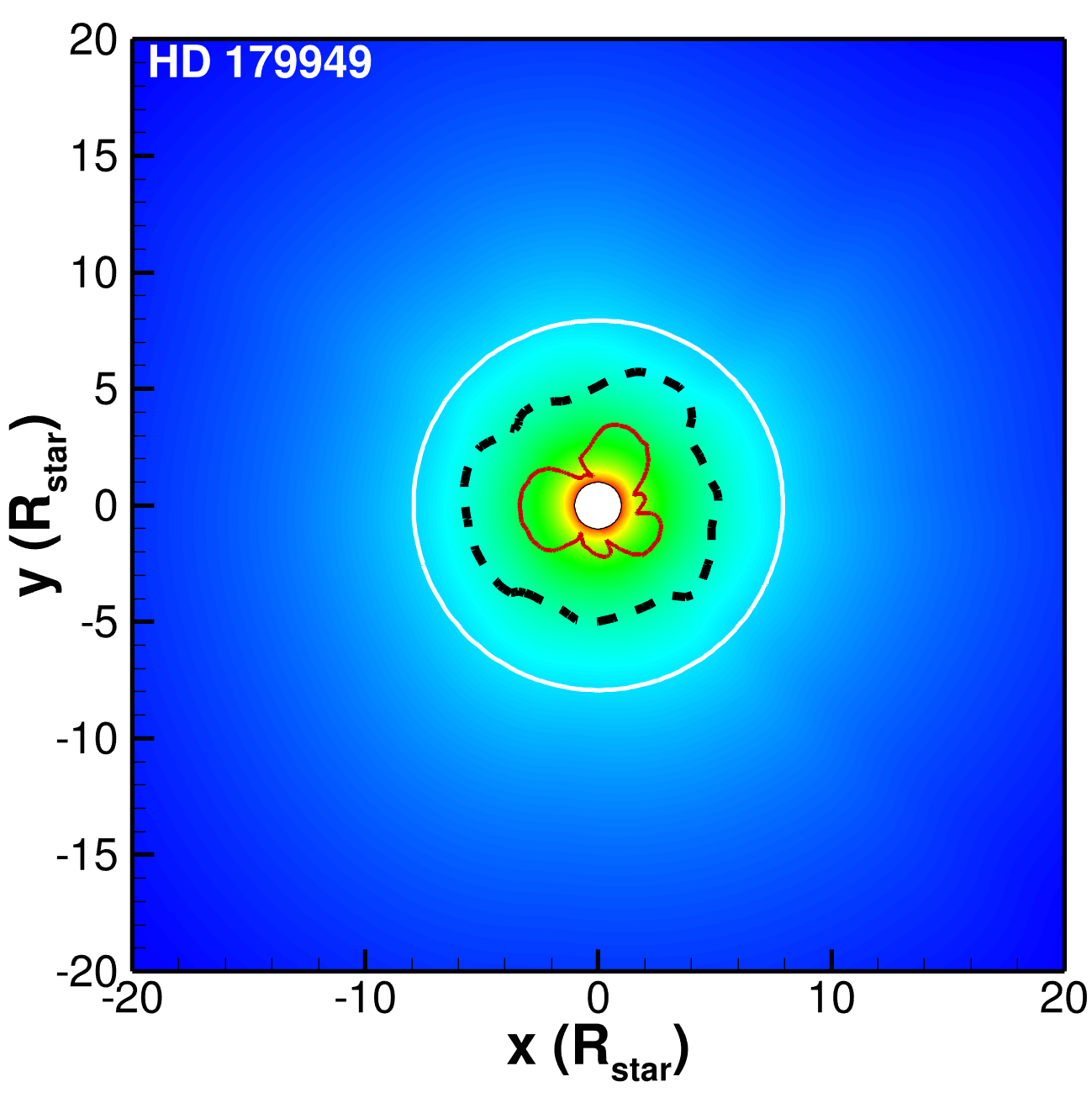}
\includegraphics[height=53mm]{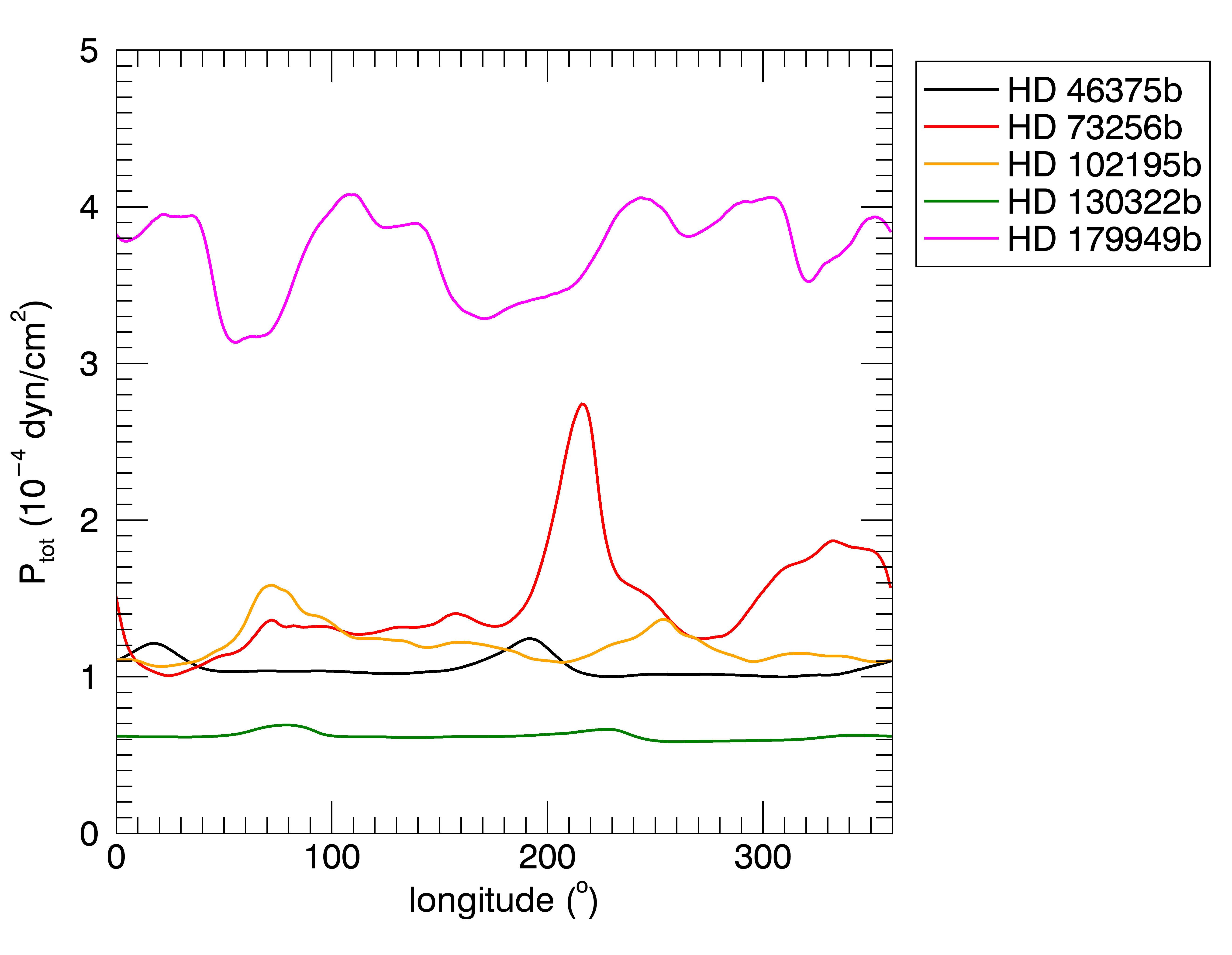}
\caption{Distribution of the total pressure $p_{\rm tot}$ experience by a planet as it orbits at the equatorial plane of each star in our simulations. The black dashed and solid red contours are cuts of the fast magnetosonic and \alf\ surfaces of the stellar wind, respectively, at the equatorial plane. The white lines indicate the orbital radii, taken to be circular, of the hot-Jupiters. The last panel shows the total local pressure at these orbits as a function of subplanetary longitude. \label{fig.ptot}}
\end{figure*}

The orbital planes of the planets considered in this work are not known. Here, we assume their orbits lie in the equatorial plane of the star. This seems to be a reasonable hypothesis for our targets (cf. Table~\ref{tab.sample}), as planets orbiting stars cooler than $6200~$K have been observed to have small (projected) obliquities \citep{2010ApJ...718L.145W}. 
 Figure~\ref{fig.ptot} shows the total pressure $p_{\rm tot}$ (i.e., the sum of thermal, magnetic and ram pressures) experienced by a planet as it orbits at the equatorial plane of the stars. Note that the ram pressure term must take into account the relative motion of the planet through the interplanetary medium. Here, we assume prograde motion of the planetary orbit relative to the stellar rotation. The white circles indicate the orbital radii of each hot-Jupiter, taken here to be circular (note that for the systems investigated here the eccentricities are rather small, $< 0.06$). The colour-bar is the same for the five images, illustrating that the total pressure varies from planet to planet. The last panel in Figure~\ref{fig.ptot} shows the total {\it local} pressure at the planetary orbits as a function of subplanetary longitude (see also Table~\ref{tab.resultsp}). For the cases studied here, at these orbital distances, the dominant term in the total pressure is the ram pressure of the relative motion of the planet through the wind. The values of the local total pressure are within $(0.58 - 4.1) \times 10^{-4}$~dyn~cm$^{-2}$, which are about 4 orders of magnitude larger than the ram pressure of the solar wind at the Earth's orbit ($1.8 \times 10^{-8}$~dyn~cm$^{-2}$, \citealt{2014A&A...570A..99S}). We also note that there is some variability in the local total pressure, showing that the planets interact with the varying environment of the star along their orbits. In the case of HD~73256, the amplitude of this variability is the highest among the cases studied here and is due to the peak (which is a factor of $1.9$ above the average value of $p_{\rm tot}$ of HD~73256) at  $\sim 220$~deg. This peak is caused by a fast wind stream, associated to the magnetic feature seen in the surface magnetograms at longitude $\sim 225~$deg (Fig.~\ref{fig.maps}). A similar feature appears in Figures~\ref{fig.rM} and \ref{fig.radio} that we present later.
  Variability on larger timescales due to intrinsic variations of the stellar magnetic field can also alter the environment surrounding planets \citep{2011MNRAS.414.1573V, 2012MNRAS.423.3285V,2013MNRAS.436.2179L}, but it is not considered in the present work. 

\subsection{Exoplanetary bow shocks: sizes and orientations}
\begin{figure}
\includegraphics[width=85mm]{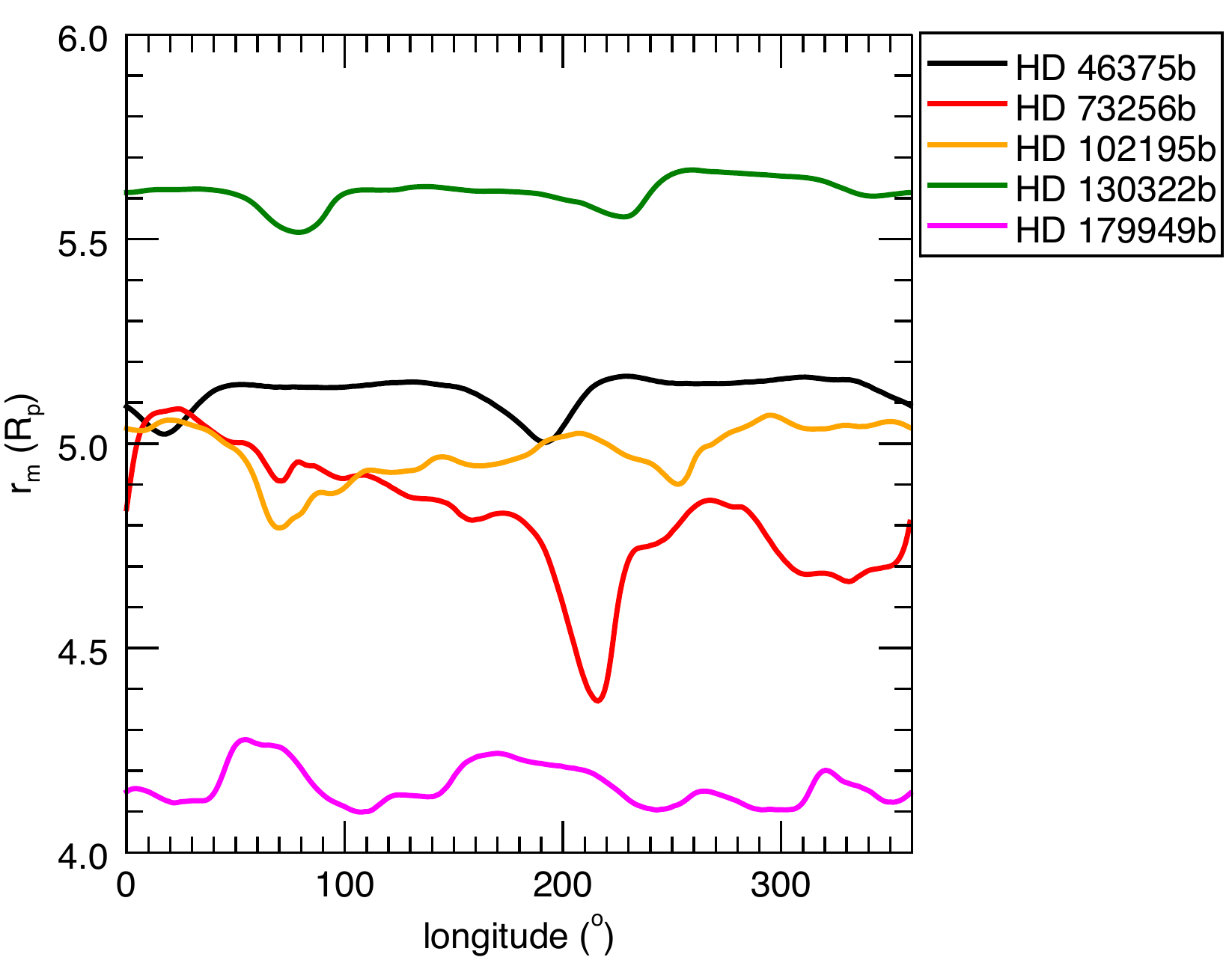}\\
\includegraphics[width=85mm]{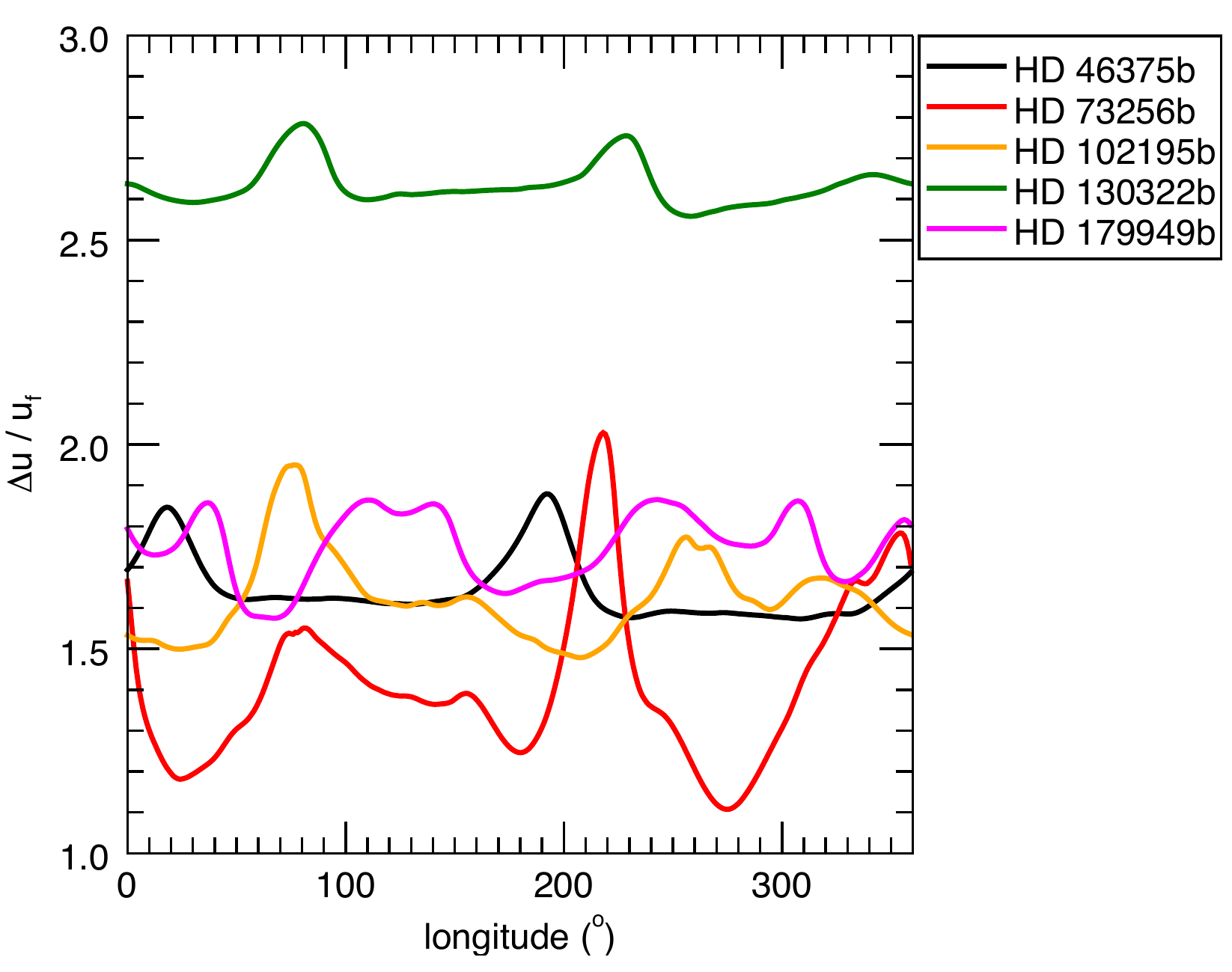}\\
\includegraphics[width=85mm]{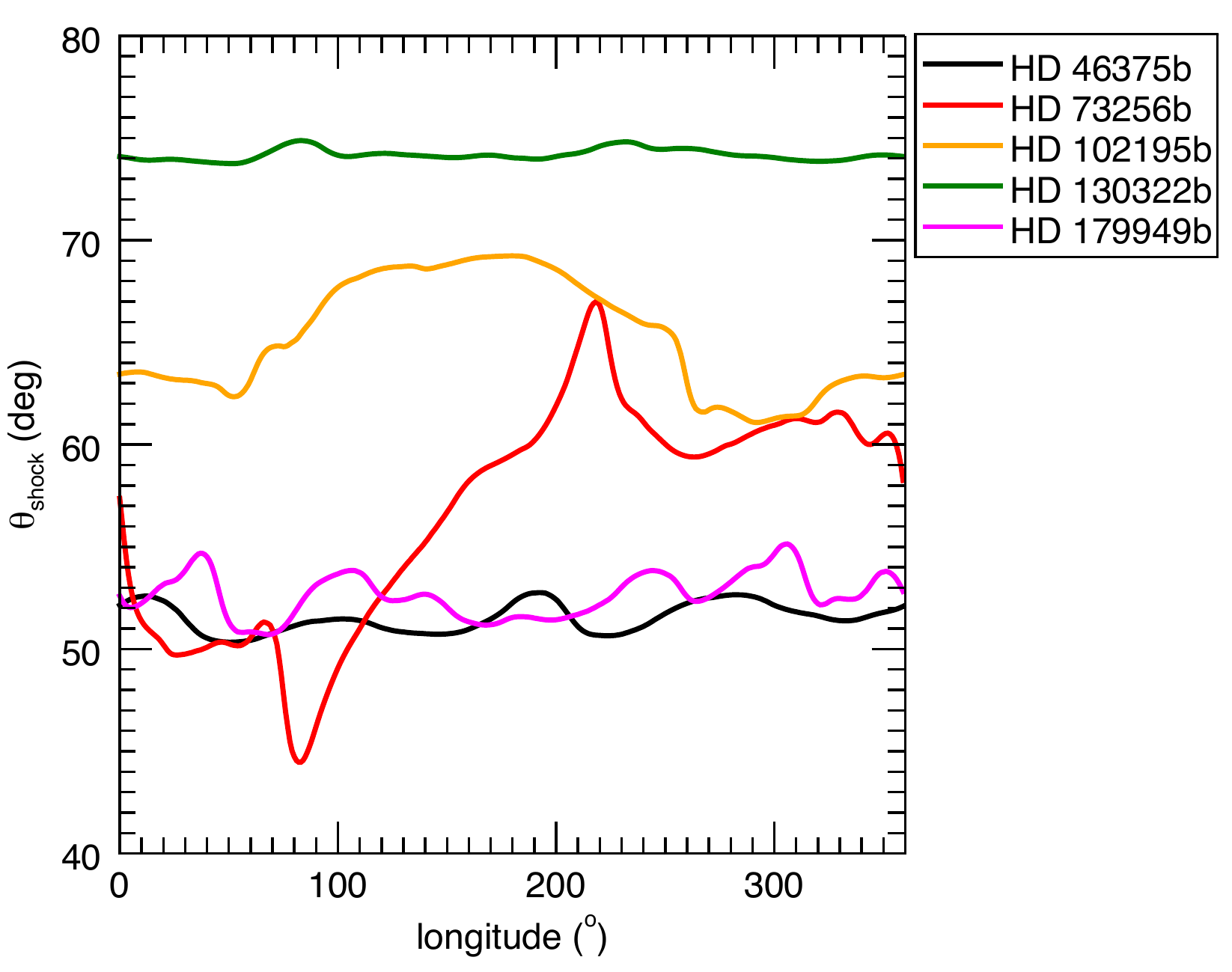}
\caption{As a magnetised planet orbits around its host star, it probes regions of the stellar wind with different properties. As a consequence, its magnetospheric size and shock orientation change. Upper panel: the magnetospheric stand-off distance for the hot-Jupiters studied here as a function of subplanetary longitude. Middle panel: the ratio between the relative velocity of the planet and the local fast magnetosonic velocity. Bottom panel: the angle formed between the shock normal and the tangent of a circular orbit. The top and bottom figures assume the hot-Jupiters have a dipolar field of $7~$G at their equator. \label{fig.rM}}
\end{figure}

If a planet is magnetised, its magnetic field can act as shield for the stellar wind, deflecting the wind particles and potentially preventing the wind from reaching down to the planetary atmosphere. A way to estimate the size of this stand-off distance is by pressure balance between the local total pressure of the interplanetary medium (i.e., the stellar wind) and the planet total pressure. Thus, at the interaction zone, we have
\begin{equation}\label{eq.equilibrium}
 p_{\rm tot} = \frac{B_{{p},r_M}^2}{8\pi} ,
\end{equation}
where $B_{{p},r_M}$ is the planetary magnetic field intensity at a distance $r_M$ from the planet centre. Eq.~(\ref{eq.equilibrium}) neglects the planetary thermal pressure component on the right side. Because of the exponential decay of planetary densities, at the height of a few planetary radii, the thermal pressure is usually negligible compared to the planetary magnetic pressure. If we assume the planetary magnetic field is dipolar, we have that $B_{{p},r_M} = B_{p, {\rm eq}} (R_p/r_M)^3$, where $R_p$ is the planetary radius and $B_{p, {\rm eq}}$ its surface magnetic field at the equator (half the value of the intensity at the magnetic pole). For a planetary dipolar axis aligned with the rotation axis of the star, the magnetospheric size of the planet is given by
\begin{equation}\label{eq.r_M}
 \frac{r_M}{R_p} = \left[ \frac{B_{p, {\rm eq}}^2}{8 \pi p_{\rm tot}} \right]^{1/6}.
\end{equation}
In the absence of observational constraints, we assume the hot-Jupiters studied here to host magnetic fields similar to Jupiter's. Figure~\ref{fig.rM}a shows the magnetospheric sizes of these hot-Jupiters assuming $B_{p, {\rm eq}}=7~$G (i.e., half of the maximum observed field of Jupiter of $\sim14~$G, \citealt{1975Sci...188..451S,1992AREPS..20..289B}.). The average estimated magnetospheric sizes range from about $\langle r_M\rangle=4.2~R_p$ for HD~179949b to $\langle r_M\rangle = 5.6~R_p$ for HD~130322b (see Table~\ref{tab.resultsp}). Variations in $r_M$ along the planetary orbit are roughly $\sim 10\%$.  This variation occurs because, as the planet goes along its orbit and probes regions with different $p_{\rm tot}$, its magnetospheric size reacts accordingly, becoming smaller when the external $p_{\rm tot}$ is larger and vice-versa. 

\begin{table*} 
\centering
\caption{Derived characteristics of the hot-Jupiters and of their local environments. The columns are, respectively, the planet name, the averages of the local velocity of the wind in the reference frame of the planet,   wind density,  magnetic field strength, wind temperature, total pressure,  planetary magnetospheric radius, shock angle, auroral oval opening angle and fractional area of the polar cap. These quantities were averaged over the subplanetary longitude. Values in brackets represent the minimum and maximum values of the averaged quantity. \label{tab.resultsp}}    
\begin{tabular}{lccccccccccccc}  
\hline

Planet	&	$			\langle \Delta u \rangle			$	&	$			\langle n \rangle			$	&	$			\langle |B| \rangle			$	&	$			\langle T \rangle			$	&	$			\langle p_{\rm tot} \rangle			$	&	$			\langle r_M\rangle			$	&	$			\langle \theta_{\rm shock} \rangle			$	&	$			\langle \alpha_0 \rangle			$	&	$			\langle A_{\rm auroral} \rangle			$	&	$			\langle \phi_{\rm radio} \rangle			$	\\
ID	&				(km s$^{-1}$)				&				(10$^5$ cm$^{-3}$)				&				(mG)				&				($10^6$ K)				&	$			(10^{-4} \frac{\rm dyn}{\rm cm^{2}})			$	&	$			(R_p)			$	&				(deg)				&				(deg)				&	$			(A_{\rm planet})			$	&				(mJy)				\\ \hline
HD46375b	&	$			234			$	&	$			1.8			$	&	$			8.8			$	&	$			0.87			$	&	$			1.1			$	&	$			5.1			$	&	$			52			$	&	$			26.2			$	&	$			0.10			$	&	$			0.037			$	\\
	&	$	[	228	,	242	]	$	&	$	[	1.7	,	2.0	]	$	&	$	[	0.55	,	11	]	$	&	$	[	0.86	,	0.91	]	$	&	$	[	1.0	,	1.2	]	$	&	$	[	5.0	,	5.2	]	$	&	$	[	50	,	53	]	$	&	$	[	26.0	,	26.6	]	$	&	$	[	0.10	,	0.11	]	$	&	$	[	0.036	,	0.043	]	$	\\
HD73256b	&	$			263			$	&	$			2.0			$	&	$			17			$	&	$			1.1			$	&	$			1.6			$	&	$			4.8			$	&	$			57			$	&	$			27.1			$	&	$			0.11			$	&	$			0.045			$	\\
	&	$	[	217	,	345	]	$	&	$	[	1.6	,	2.6	]	$	&	$	[	2.6	,	26	]	$	&	$	[	0.91	,	1.6	]	$	&	$	[	1.0	,	2.7	]	$	&	$	[	4.4	,	5.2	]	$	&	$	[	44	,	67	]	$	&	$	[	26.1	,	28.5	]	$	&	$	[	0.10	,	0.12	]	$	&	$	[	0.027	,	0.081	]	$	\\
HD102195b	&	$			288			$	&	$			1.5			$	&	$			14			$	&	$			0.96			$	&	$			1.3			$	&	$			5.0			$	&	$			65			$	&	$			26.6			$	&	$			0.11			$	&	$			0.067			$	\\
	&	$	[	240	,	338	]	$	&	$	[	1.1	,	2.0	]	$	&	$	[	3.6	,	18	]	$	&	$	[	0.87	,	1.2	]	$	&	$	[	1.1	,	1.6	]	$	&	$	[	4.8	,	5.1	]	$	&	$	[	61	,	69	]	$	&	$	[	26.2	,	27.1	]	$	&	$	[	0.10	,	0.11	]	$	&	$	[	0.054	,	0.086	]	$	\\
HD130322b	&	$			322			$	&	$			0.6			$	&	$			2.3			$	&	$			0.78			$	&	$			0.62			$	&	$			5.6			$	&	$			74			$	&	$			25.0			$	&	$			0.09			$	&	$			0.055			$	\\
	&	$	[	316	,	334	]	$	&	$	[	0.6	,	0.7	]	$	&	$	[	0.36	,	2.9	]	$	&	$	[	0.77	,	0.79	]	$	&	$	[	0.58	,	0.69	]	$	&	$	[	5.5	,	5.7	]	$	&	$	[	74	,	75	]	$	&	$	[	24.8	,	25.2	]	$	&	$	[	0.09	,	0.10	]	$	&	$	[	0.053	,	0.061	]	$	\\
HD179949b	&	$			243			$	&	$			5.9			$	&	$			9.6			$	&	$			0.97			$	&	$			3.8			$	&	$			4.2			$	&	$			53			$	&	$			29.3			$	&	$			0.13			$	&	$			0.112			$	\\
	&	$	[	225	,	257	]	$	&	$	[	5.5	,	6.2	]	$	&	$	[	1.4	,	15	]	$	&	$	[	0.96	,	0.99	]	$	&	$	[	3.1	,	4.1	]	$	&	$	[	4.1	,	4.3	]	$	&	$	[	51	,	55	]	$	&	$	[	28.9	,	29.6	]	$	&	$	[	0.12	,	0.13	]	$	&	$	[	0.092	,	0.127	]	$	\\ \hline

\end{tabular}
\end{table*}

Over-plotted to Figure~\ref{fig.ptot} are the contours  at the equatorial plane of the \alf\ surface (red lines) and the magnetosonic surface (black lines) of the stellar wind. In all the cases studied here, the planets orbit at regions of super fast magnetosonic velocities of the stellar wind. One exception is the case of HD~73256b, in which a small part of its orbit (white circle) lies within the fast magnetosonic surface of the wind. This does not necessarily mean that at these orbital positions a bow shock will not be formed surrounding HD~73256b's magnetosphere. Rather, it is the relative velocity of the planet orbiting through the stellar wind 
\begin{equation}
\Delta {\bf u} = {\bf u}-u_K \boldsymbol{\hat{\varphi}},
\end{equation}
where $u_K$ is the (purely azimuthal) Keplerian velocity of the planet, that should be compared to the fast magnetosonic velocity of the local plasma $v_f = (c_s^2 + v_A^2)^{1/2}$, where $c_s$ is the local sound speed. Figure~\ref{fig.rM}b shows the fast magnetosonic Mach number ($\Delta u/v_f$) calculated at the orbital radii of the hot-Jupiters, where we see that the relative planetary velocity is always super-fast magnetosonic (i.e., $\Delta u/v_f>1$), indicating that the magnetosphere of these planets are surrounded by bow shocks.

It has been proposed that these bow shocks might absorb at specific wavelengths, generating asymmetric transit lightcurves \citep{2010ApJ...722L.168V}. This is particularly relevant for the case of hot-Jupiters, in which the orientation of the bow shock is shifted towards the direction of planetary motion (as opposed to the bow shocks surrounding the solar system planets, which are largely formed facing the Sun). These `sideways' bow-shocks present the best conditions for detection during planetary transits \citep{2011MNRAS.416L..41L,2013MNRAS.436.2179L}. Although the hot-Jupiters investigated here are not transiting and do not have constrained orbital inclinations, there has been cases in the literature of non-transiting (but grazing) exoplanets whose extended atmospheres might undergo partial transit \citep{2012A&A...547A..18E}. Likewise, combined with the orbital inclinations, it is possible that bow shocks of non-transiting planets might be visible if they graze the stellar disc. We, here, do not model the 3D extent of bow shocks, as done in \citet{2011MNRAS.416L..41L,2013MNRAS.436.2179L}, but we can calculate the angle between the shock normal and the tangent of a circular orbit  
\begin{equation}
\theta_{\rm shock} = \arctan \left( \frac{u_r}{|u_K-u_\varphi|}\right)
\end{equation}
\citep{2010ApJ...722L.168V}. Along its orbital path, the planet probes regions of the wind with different velocities, which implies that the orientation of the bow shock that forms surrounding planetary magnetospheres changes along the planetary orbit. This can be seen in Figure~\ref{fig.rM}c and Table~\ref{tab.resultsp}, where we present $\theta_{\rm shock}$ as a function of the subplanetary longitude. We present in Table~\ref{tab.resultsp} the average shock angle $\langle \theta_{\rm shock} \rangle$ of the bow shock of each of these hot-Jupiters, where we note that they range from about $52^{\rm o}$ to $74^{\rm o}$.
 
\subsection{Exoplanetary auroral ovals: escape channels and radio emission}\label{sec.polarflows}
As the planetary magnetosphere extent is reduced, the size of the `auroral oval', which is the amount of planetary area with open magnetic field lines, increases. Along these open field lines, particles can be transported to/from the interplanetary space, affecting, for instance, the amount of atmospheric mass loss \citep{2011ApJ...730...27A}. We estimate here the size of the auroral region of the planet as follows. Assuming the planet to have a dipolar magnetic field, aligned with the planetary orbital spin axis, the colatitude of the largest closed field line of the planet, which defines the boundary between open- and closed-field line regions, can be estimated as $\alpha_0=\arcsin ({R_p}/{r_M})^{1/2}$ \citep{1975JGR....80.4675S,2010Sci...327.1238T}. This implies in a fractional area of the planetary surface that has open magnetic field lines 
\begin{equation}\label{eq.area}
\frac{A_{\rm polar~cap}}{A_{\rm planet}} = (1-\cos \alpha_0),
\end{equation} 
 \citep{2013A&A...557A..67V}. Therefore, in addition to making $r_M$ smaller, a stronger external pressure of the stellar wind exposes a larger area of the polar cap of the planet. Table~\ref{tab.resultsp} shows the average, minimum and maximum angles of the auroral ovals $\langle \alpha_0 \rangle$ and fraction of open area $\langle A_{\rm polar~cap} \rangle$ as calculated by Eq.~(\ref{eq.area}). For the hot-Jupiters analysed here, $\langle \alpha_0 \rangle$ ranges between $25^{\rm o}$ and $29^{\rm o}$, and $\langle A_{\rm polar~cap} \rangle$ ranges between $9\%$ and $13\%$.  For comparison, the size of the auroral oval in Saturn is $\alpha_0 \simeq 10^{\rm o}$ -- $20^{\rm o}$  \citep{2005Natur.433..717C} and at the Earth it is $\alpha_0 \simeq 17^{\rm o}$ -- $20^{\rm o}$ \citep{2009AnGeo..27.2913M}. Using Eq.~(\ref{eq.area}), a rough estimate indicates that the open-field-line region covers $\sim 1.5\%$ -- $6 \%$ of Saturn's surface and $\sim 4.5\%$ -- $6 \%$ of Earth's surface. This is a factor of $\sim 2$ smaller than the values we derive for the hot-Jupiters in our sample, but not as extreme as the cases of planets orbiting at the habitable zone of more active stars \citep{2013A&A...557A..67V}.
 
Planetary radio emission takes place in a hollow cone of half-aperture angle given by the auroral oval co-latitude $\alpha_0$. It has been recognised that the radio emission of the Earth and the four giant planets of the solar system correlates to the local characteristics of the solar wind  \citep[e.g.,][]{1998JGR...10320159Z}, an indication that radio emission is powered by the local solar wind. Analogously, it is expected that when exoplanets interact with the wind of their host stars, they would also be sources of radio emission.

We use the results of our stellar wind simulations to calculate the kinetic power of the wind, at the orbital radii of the hot-Jupiters studied here. Our approach follows closely the one in  \citet{2012MNRAS.423.3285V}. The kinetic power $P_k$ of the impacting wind on the planet is approximated as the ram pressure of the particles $\rho (\Delta u)^2$ impacting on the planet, with effective cross-section $\pi r_M^2$, at a relative velocity $\Delta {\bf u}$
\begin{equation}\label{eq.pK}
P_k \simeq \rho (\Delta u)^3 \pi r_M^2 .
\end{equation}
The radio flux can be written as
\begin{equation}\label{eq.radioflux}
\phi_{\rm radio}  = \frac{P_{\rm radio}}{d^2 \omega \Delta f}  = \frac{\eta_k P_{\rm k}}{d^2 \omega \Delta f} 
\end{equation}
where $d$ is the distance to the system, $\omega = 2\times 2 \pi (1 - \cos \alpha_0)$ is the solid angle of the hollow emission cone (defined by the auroral oval), and $\Delta f$ is the frequency of emission. In the last equality, we assumed a linear efficiency $\eta_k$ in converting the power released from the dissipation of kinetic wind energy to radio emission (`radiometric Bode's law'). We adopt $\eta_k  = 10^{-5}$, as derived from observations of the Solar System planets \citep{2007P&SS...55..598Z}. Here, we assume that the emission bandwidth $\Delta f$ is approximately the cyclotron frequency \citep{2007P&SS...55..618G}:
\begin{equation}\label{eq.fcyc}
\Delta f =  \frac{e B_p(\alpha_0) }{m_e c} = 2.8 \left[ \frac{B_p(\alpha_0)}{1~{\rm G}}\right] ~{\rm MHz} ,
\end{equation}
where $m_e$ is the electron mass and $c$ the speed of light. $B_p(\alpha_0)$ is the planet's magnetic field strength at colatitude $\alpha_0$ of the auroral ring. For a dipolar field, $B_p(\alpha_0)= B_{p, {\rm eq}} (1 + 3 \cos \alpha_0)^{1/2}$.

To compute the radio flux (Eq.~\ref{eq.radioflux}), we need to know the physical size of $r_M$.  
This value, normalised to the planet's radius, is given in Figure~\ref{fig.rM}a and we further assume planetary radii of $1.5R_{\rm Jup}$ for all the hot-Jupiters analysed in this work (note that they are non-transiting planets and therefore do not have observationally-determined radii). Eq.~(\ref{eq.radioflux}) is the only place where the physical size of the exoplanet is required and different choices of $R_p$ influence the estimated radio flux as $\phi_{\rm radio} \propto r_M^2 \propto R_p^2$.

\begin{figure}
\includegraphics[width=85mm]{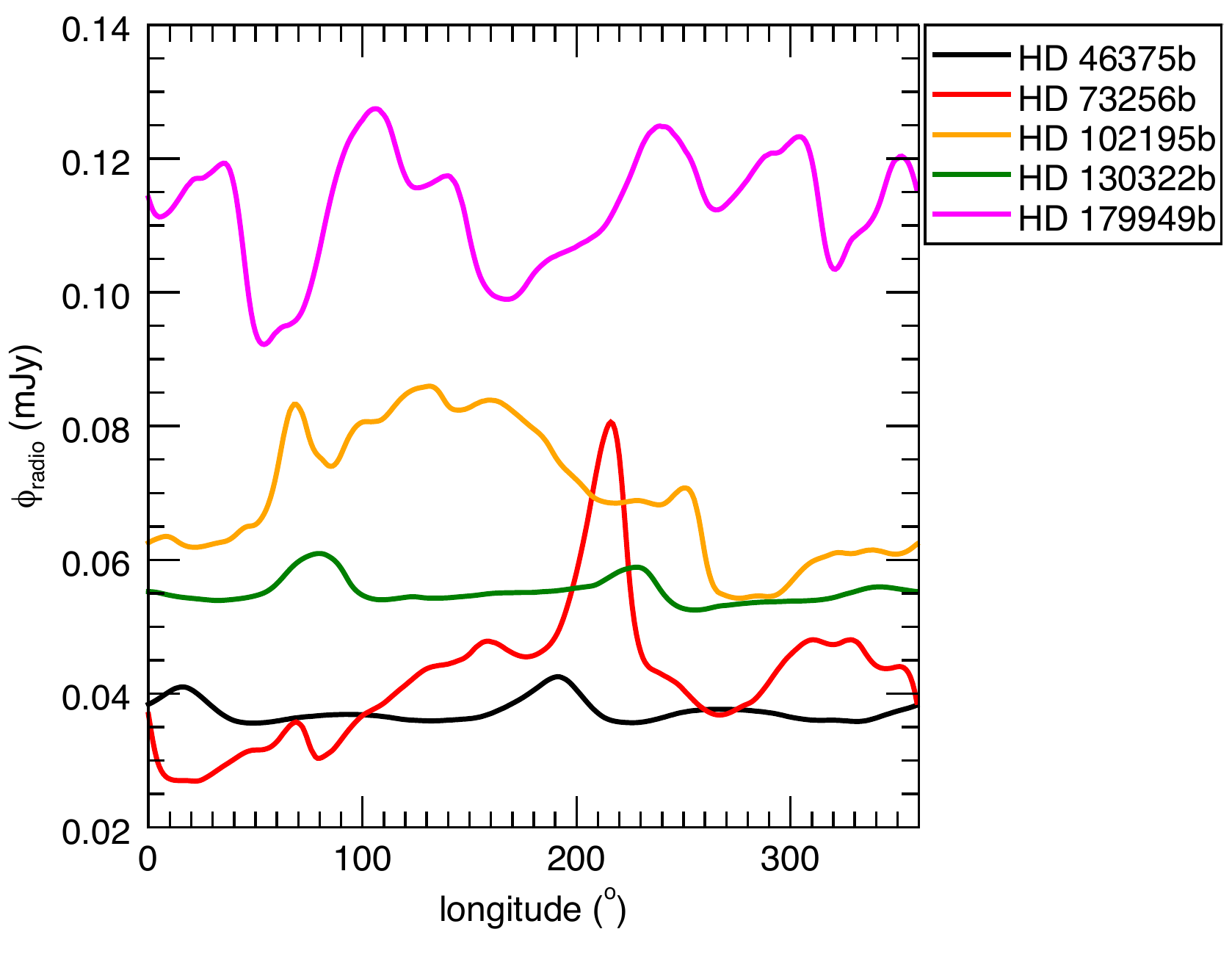}\\
\includegraphics[width=85mm]{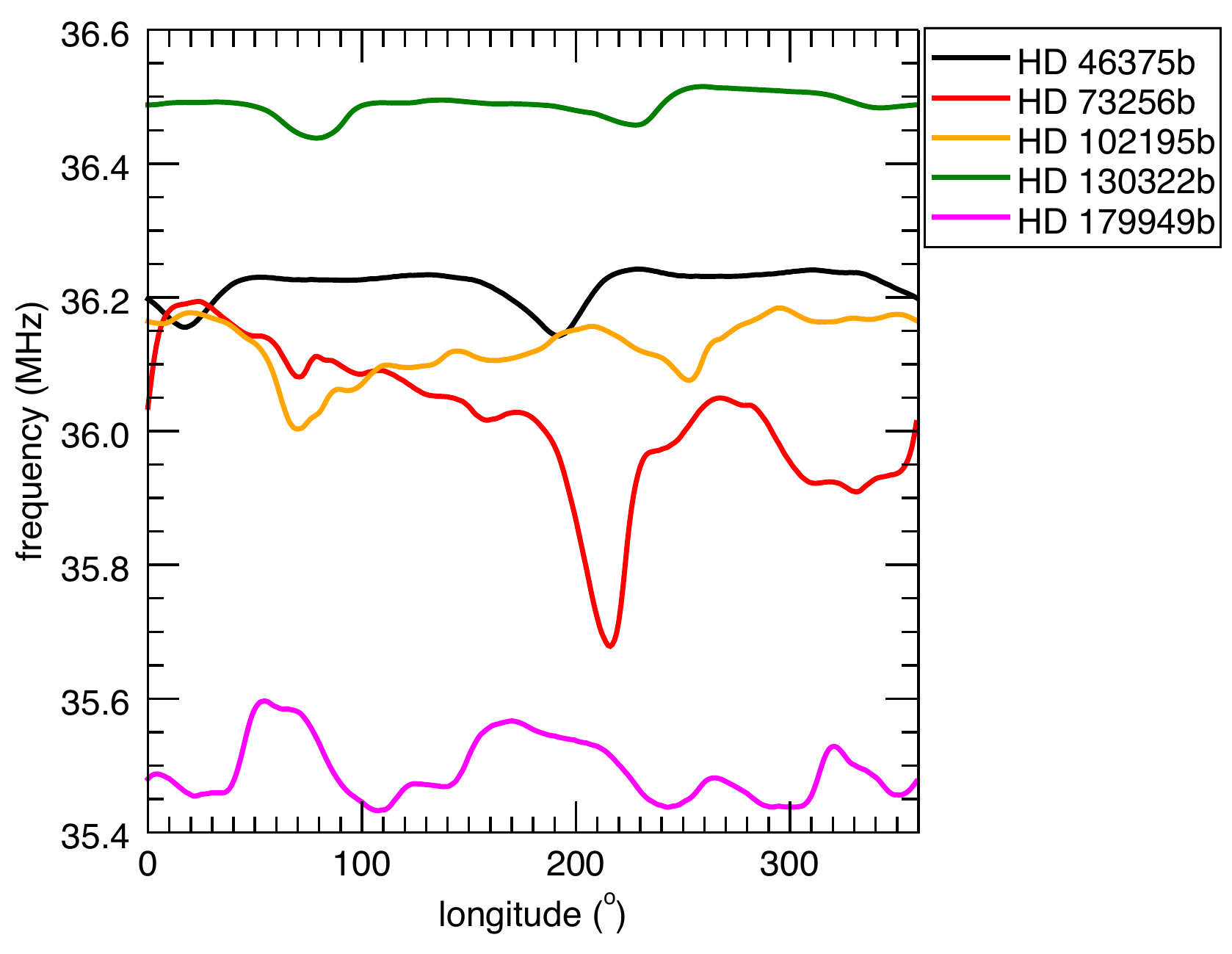}
\caption{{The predicted radio flux (Eq.~\ref{eq.radioflux}) computed using the results of our wind simulations (top) and associated frequency of emission (bottom) assuming the emission bandwidth is the cyclotron frequency (Eq.~\ref{eq.fcyc}). These results assume a dipolar exoplanetary magnetic field, whose intensity is $7~$G at the equator.} \label{fig.radio}}
\end{figure}

Figure~\ref{fig.radio}a shows the radio flux computed using the results of our wind simulations and Figure~\ref{fig.radio}b shows the calculated frequency of emission. We find that the predicted emission frequency occurs at $\sim 36$~MHz and the radio fluxes range between $0.02$ and $0.13$~mJy among all the cases studied here (see also Table~\ref{tab.resultsp}). Values of radio fluxes such as these (including the peak values that occur at favourable phases) should be challenging to be observed with present-day technology, such as with LOFAR, whose sensitivity at $20$ to $40$~MHz is $\gtrsim30$ to $3$mJy, respectively, for a one-hour integration time \citep{2011RaSc...46.0F09G}. It is likely, however, that even these small radio fluxes will be detectable with future higher sensitivity arrays, such as the SKA-low array system. 

Among the systems studied here, HD~179949b has the highest estimated radio flux. This occurs for two reasons. First, this exoplanet has the closest orbital radius and, because of that, $\rho \Delta u^3$ is the largest among our sample; for the same reason, it also has the smallest $r_M$ (cf.~Tables \ref{tab.sample} and \ref{tab.resultsp}). In spite of the smallest cross-section $\pi r_M^2$, the large $\rho \Delta u^3$ term is more important in Eq.~(\ref{eq.pK}), which results in the largest stellar wind kinetic power impacting on the magnetosphere of the exoplanets studied here. Second, the closest distance to the HD~179949 system also favours a larger radio flux (Eq.~(\ref{eq.radioflux})). 

It is also worth comparing the emission calculated here and the values calculated for $\tau$~Boo~b and HD~189733b\footnote{Note that the stellar wind simulations presented in \citet{2012MNRAS.423.3285V} and \citet{2013MNRAS.436.2179L} have the same assumptions as the ones shown in the present work.}. Using the same radio emission model presented here, \citet{2012MNRAS.423.3285V} estimated the radio flux of $\tau$~Boo~b at different epochs of the host star's magnetic cycle. They found the radio flux of $\tau$~Boo~b to be of the order of $0.5$ -- $0.9$~mJy. We can also use the simulations presented in \citet{2013MNRAS.436.2179L} to compute the radio flux of HD~189733b. Assuming a planetary radius of $R_p = 1.15~R_{\rm Jup}$ and a distance of $19.3~$pc, we calculate the radio flux of HD~189733b to be on average $0.47$~mJy (peak at 0.98~mJy) for the case where the observed stellar magnetic map is derived from the 2007~June observations and $0.23~$mJy (peak at $0.51$~mJy) for the 2008~July map (cf.~\citealt{2010MNRAS.406..409F}).

The radio fluxes computed for $\tau$~Boo~b and HD~189733b are therefore considerably larger than the values computed for the exoplanets presented here, having better prospects for being detected. The reason why these two systems have higher radio fluxes is similar to the reasons discussed for the case of HD~179949b: a combination of closer orbital radii ($6.8~R_\star$ and $8.6~R_\star$ for $\tau$~Boo~b and HD~189733b, respectively) and closer distances to the systems ($15.6$ and $19.3$~pc). It is also expected that exoplanets orbiting young stars (with denser stellar winds) are likely to produce higher radio fluxes \citep{2005A&A...437..717G,2010ApJ...720.1262V}, presenting also better prospects for detection of exoplanetary radio emission.

Radio fluxes estimated for the $5$ hot Jupiters studied here have been estimated by other authors. For instance, \citet{2011RaSc...46.0F09G} predicted radio fluxes that are larger than the values predicted here by a factor of $500$ -- $2000$ (compared to the case for their rotation-independent planetary magnetic field model). Although our radio emission model is similar to the one used in \citet{2011RaSc...46.0F09G} (i.e., both our models assume a `radiometric Bode's law', in which a fraction of the dissipation of the wind power is converted into planetary radio emission), we attribute the difference found between their work and the present one due to the different models assumed for the stellar wind and stellar magnetic field, as well as for the assumed planetary magnetic field intensities. For the stellar wind, \citeauthor{2011RaSc...46.0F09G}'s work assumes a spherically symmetric, isothermal wind model \citep{1958ApJ...128..664P}. The velocity and density structures are scaled with respect to the age of the system, based on the age relations found by \citet{1980asfr.symp..293N} and \citet{2005ApJ...628L.143W}. For the planetary magnetic field, \citeauthor{2011RaSc...46.0F09G}'s work assumes either a case where the planetary dynamo is independent  \citep{2010A&A...522A..13R} or dependent \citep{2004A&A...425..753G} on the planetary rotation. \citet{2011RaSc...46.0F09G} showed that the intensity of the planetary magnetic field affects the frequency of the emission (as in our model) and that the radio flux has a strong dependence with the intensity of the planetary magnetic field (contrary to our model). More recently, \citet{see2015} studied the variability of exoplanetary radio emission for a sample of planet host stars, which includes the objects studied in the present work. Similar to our model, their model incorporates the realistic large-scale geometry of the stellar magnetic field, but their radio emission model differs from ours. Instead, their study was based on the model developed by \citet{2008A&A...490..843J}, which computes the radio emission generated by energetic electrons released in the reconnection  between stellar and exoplanetary magnetic field lines, without assuming the a priori relation of the `radiometric Bode's law'. Despite the differences in these models, the radio fluxes estimated by \citet{see2015} and by us are very similar, within a factor of $1$ -- $4$, except for HD130322, in which we estimate radio fluxes that are $100$ times larger than theirs. In addition to providing information on exoplanetary's magnetic field, detection of exoplanetary radio emission would clearly provide invaluable constraints to stellar wind models as well. 

\section{Discussion}\label{sec.discussion}
\subsection{Limitations of the models}\label{sec.limitations}
The stellar wind models presented in this paper use as input the observationally reconstructed stellar magnetic field and are, therefore, more realistic (and provide an advance) compared to models that are non-magnetised or that assume simplified stellar magnetic field topologies. In spite of that, our wind models share the limitations of global, polytropic wind models. In particular, these types of models have three parameters that are poorly constrained by observations, namely, the wind base density and temperature and the temperature profile (i.e., the profile of energy deposition through the parameter $\gamma$). In this work, we have chosen to set all these three parameters to be the same for all the stars in our sample. On the other hand, parameters such as the stellar mass, radius, rotation period and magnetic field differ for each object and are constrained to values observationally-derived for each stars (Table~\ref{tab.sample}). 

\citet{Johnstone2015} recently showed that the average temperature of X-ray coronae $\langle T_{\rm cor} \rangle$ is a weak function of  X-ray flux $F_X$: $\langle T_{\rm cor} [{\rm MK}]  \rangle= 0.11 (F_X/[{\rm erg~s}^{-1}{\rm cm}^{-2}])^{0.26}$ (see also \citealt{2005ApJ...622..653T}). Using their relation, the X-ray luminosities compiled in \citet{2014MNRAS.441.2361V} and the stellar radii from Table~\ref{tab.sample}, we estimate $\langle T_{\rm cor} \rangle$ to be in the range between $2$ and $3.6$~MK for the stars in our sample. Naively, one could expect that the temperature at the base of the wind is related to the temperature of the closed X-ray corona (and this is the case for our Sun), but it is not clear if this relation is true for other stars. Therefore, in the absence of a stronger constraint, in our models, we adopt a wind base temperature of $2$~MK, typical of stellar coronae of solar-type stars. We adopt $\gamma$ that is the same as the effective adiabatic index measured in the solar wind \citep{2011ApJ...727L..32V}. For the base density, we adopted a value of $10^9 {\rm cm}^{-3}$. Ideally, observations of mass-loss rates of cool dwarf stars would allow us to place better constraints on the densities. However the lack of, or difficult-to-obtain, observational signatures of these winds make constraints of base density (or mass-loss rates) challenging to be obtained.

To investigate how our results change with the change in the wind base density, we performed a stellar wind simulation of HD~46375 that results in a mass-loss rate ($\mdot=2.9 \times 10^{-14}~\msano$) that is similar to the one observed in the solar wind ($\mdot=2 \times 10^{-14}~\msano$). Compared to the values of HD~46375 reported in Table~\ref{tab.results}, in this simulation, we found a mass-loss rate that is a factor of $6.5$ smaller, $\jdot$ that is a factor $3$ smaller and $\Phi_0$ that is a factor $1.3$ smaller. 

Locally, the hot-Jupiter HD~46375b experience a total external pressure whose average value (averaged over the longitude of the subplanetary point) is a factor of $5.6$ smaller than the value presented in Table~\ref{tab.resultsp}. Because $r_M$ is weakly dependent on $p_{\rm tot}$ ($r_M \propto p_{\rm tot}^{-1/6}$), the value of $r_M$ we estimated before is smaller by a factor of only $1.3$. In spite of the larger the cross-section of the planetary magnetosphere, the radio flux decreased by a factor of $2.3$, caused by the decrease in the ram pressure [Eq.~(\ref{eq.pK})].

Another parameter we have assumed in our models is the planetary magnetic field intensity. As discussed in Section~\ref{sec.introBp}, this is a quantity that has yet not been measured in exoplanets. Here, we adopted a magnetic field intensity which is similar to the value of Jupiter. We can also estimate how the magnetospheric size we presented in Fig.~\ref{fig.rM}a would have changed if a different field strength were to be adopted. Because $r_M \propto B_p^{1/3}$, a strength that is a factor of $2$ smaller would decrease the reported values of $r_M$ by $2^{1/3}$ (i.e., only $25\%$). In spite of that, this would not have significantly altered the computed radio flux (our radio flux model is weakly dependent on the planetary field strength; see discussion in \citealt{2012MNRAS.423.3285V}), but would have decreased the frequency of the emission by a factor of $2$, making it not possible to be observed from the ground, due to the Earth's ionospheric cut-off in frequencies. Indeed, one of the possibilities that exoplanetary radio emission has not been detected so far might be due to a frequency mismatch between the emission source and that of the search instruments \citep{2000ApJ...545.1058B}.

\subsection{Exoplanetary system conditions for detectability at radio wavelengths}
Because of the cyclotron nature of the magnetospheric radio emission, exoplanets with higher magnetic field strengths emit at higher frequencies, where the detection sensitivity is larger.  For instance, an exoplanet with a magnetic field of about $40$ -- $50$~G emits at the frequency range between $110$ and $140$ MHz. The sensitivity of LOFAR at $100$ to $200$~MHz is roughly about $0.05$~mJy (cf.~Fig.~1 in  \citealt{2011RaSc...46.0F09G}). This indicates that, except for HD~46375b, all the remaining exoplanets studied here could in principle be detectable with LOFAR, if their magnetic field strengths were about $40$ -- $50$~G. Compared to Jupiter' maximum field intensity, these field strengths are about $\sim 3$ times higher. 

We can also estimate what would be the required dissipated stellar wind power to generate detectable radio signatures from the exoplanets studied here. In this exercise, we take the same exoplanetary magnetic field assumed in Section~\ref{sec.planets} (i.e., $B_{p, {\rm eq}}=7~$G). With such a magnetic field intensity, the frequency of emission is around $36$~MHz, where the LOFAR sensitivity for a one-hour integration time is about a few mJy. The radio power calculated in Section~\ref{sec.planets} yielded values of about ($1.6$ -- $5.6) \times 10^{25}$ erg~s$^{-1}$. For a radio flux of a few mJy, the required radio power of the exoplanets studied here should be higher, on the range of  ($1.1$ -- $2.1) \times 10^{27}$ erg~s$^{-1}$. To have a radio power (or, equivalently, a wind kinetic power) that is roughly 2 orders of magnitude larger, the stellar wind characteristics need to change -- either by increasing the density of the stellar wind or its velocity or both, as demonstrated next. 

From equations (\ref{eq.r_M}), (\ref{eq.pK}) and (\ref{eq.radioflux}), and assuming a ram pressure-dominated wind, one can show that 
\begin{equation}
\rho^2 \Delta u^7 \sim \frac{8 P_k^3}{\pi^2 R_p^6 B_p^2}, 
\end{equation}
such that the ratio between the values required for a radio flux of about 3~mJy to the ratio of the values calculated at Section~\ref{sec.planets} is 
\begin{equation}\label{eq.estimate}
\frac{[\rho^2 \Delta u^7]_{(3 {\rm mJy})}}{[\rho^2 \Delta u^7]_{\rm (Sect.~5)} } \sim \left( \frac{P_{k(3 {\rm mJy})}}{P_{k (\rm Sect.~5)} }\right)^3 \sim (0.5 ~{\rm to~} 3.3)\times10^5.
\end{equation}
A very crude estimate\footnote{Note that this approach is not a self-consistent one, because in this scenario the structure of the wind temperature, velocity and magnetic fields are not modified (i.e., we are assuming they remain unchanged as the structures derived in Section~\ref{sec.results}). In the self-consistent approach, if either the density of the wind or its velocity are modified, one needs to solve the coupled MHD equations to derive all the remaining quantities of the wind. However, this back-of-the-envelope calculation can give a rough estimate of how larger should the stellar wind power be in order for the radio emission to reach values above the sensitivity limit of a couple of mJy.} then tells us that either the wind density needs to increase by a factor of at least $\sim 300$ to $600$ (i.e., the square root of the values derived in Eq.~(\ref{eq.estimate})) or the velocity requires an increase of a factor of at least $\sim 5$ to $7$ (i.e., the 7-th root of the values in Eq.~(\ref{eq.estimate})). Or, alternatively, density and velocity should both change such that they obey the relation (\ref{eq.estimate}).

From Table~\ref{tab.resultsp}, a $5$ to $7$ times increase in the wind velocity implies a relative velocity $\gtrsim 1200$~km/s, which is $50\%$ larger than the speed of the fast solar wind and 3 times larger than the slow solar wind speed. In terms of density, an increase of $\sim 300$ to $600$ roughly implies a similar increase in mass loss rates and, from Table~\ref{tab.results}, this would result in $\mdot$ of at least $(2.9$ -- $24) \times10^4$ times the solar wind mass-loss rates. Such mass loss-rates are typical of very young stars, indicating that exoplanets orbiting young Suns are more likely to produce detectable levels of radio fluxes \citep{2005A&A...437..717G,2010ApJ...720.1262V}.

\section{Summary and conclusions}\label{sec.conclusions}
In this work we have investigated the interplanetary media surrounding five hot-Jupiters, namely: HD~46375b, HD~73256b, HD~102195b, HD~130322b, HD~179949b.  For that, we carried out 3D MHD stellar wind simulations, which incorporate as boundary conditions the surface magnetic field of the star reconstructed by \citet{2012MNRAS.423.1006F,2013MNRAS.435.1451F} using the Zeeman Doppler Imaging technique. The global characteristics of our wind models are presented in Table~\ref{tab.results}.

We then calculated the {\it local} characteristics of the stellar winds at the orbital radius of the hot-Jupiters, in order to characterise the interplanetary medium surrounding these exoplanets. In particular, we calculated the total pressure of the interplanetary medium  and estimated what would be the size of planetary magnetospheres in case these hot-Jupiters had a magnetic field similar to Jupiter's field. We found that magnetospheric sizes range between $4.1$ and $5.6~R_p$ and that they can vary by a few percent due to variations in the external environment of the planets, as they orbit around their parent stars. 
We also demonstrated that these planets orbits are super fast magnetosonic, indicating that bow shocks should be formed around their magnetospheres. The bow shock orientations (i.e., the angle between the shock normal and the tangent of the circular orbit) are of intermediate type, in which the shock forms at intermediate angles from the one of a shock facing the motion of the planet (`ahead shock') and the one connecting the star-planet centres (`dayside shock').

We also calculated the size of the auroral ovals of these planets. Inside these ovals, the planetary magnetic field lines are open, through which particles from the star and from the cosmos can penetrate as well as planetary atmospheric particles can escape through polar flows. On average, the auroral ovals we calculated have a half-opening angle of about $25^{\rm o}$ to $29^{\rm o}$, leaving exposed about $9\%$ to $13\%$ of the planetary area, which is a factor of $\sim 2$ larger than estimates for the Earth's and Saturn's auroral caps. Finally, we estimated the radio flux of these planets, using the analogy observed in the solar system, in which the radio emission from the magnetised planets is correlated to the solar wind power. We found small radio fluxes ranging from $0.02$ to $0.13$~mJy, which should represent a challenge to be observed with present-day technology (e.g., LOFAR; \citealt{2011RaSc...46.0F09G}), but could be detectable with higher sensitivity arrays, such as the SKA-low array system. Radio emission from  systems having closer hot-Jupiters, such as from $\tau$~Boo~b (radio flux of the order of $0.5$ -- $0.9$~mJy, \citealt{2012MNRAS.423.3285V}), HD~189733~b ($0.5$ -- $1$~mJy, calculated using the simulations from \citealt{2013MNRAS.436.2179L} and the same model as presented here), or from nearby planetary systems orbiting young stars \citep{2005A&A...437..717G,2010ApJ...720.1262V}, are likely to have higher radio fluxes, presenting thus better prospects for detection of exoplanetary radio emission.

\section*{Acknowledgements}
AAV acknowledges support from the Swiss National Science Foundation through an Ambizione Fellowship. RF acknowledges support from a STFC grant. The results of this work are based on observations acquired at
CFHT/ESPaDOnS and TBL/NARVAL. This work was carried out using the BATS-R-US tools developed at The University of Michigan Center for Space Environment Modeling (CSEM) and made available through the NASA Community Coordinated Modeling Center (CCMC). This work was supported by a grant from the Swiss National Supercomputing Centre (CSCS) under project ID s516. This work used the DiRAC Data Analytic system at the University of Cambridge, operated by the University of Cambridge High Performance Computing Service on behalf of the STFC DiRAC HPC Facility (www.dirac.ac.uk). This equipment was funded by BIS National E-infrastructure capital grant (ST/K001590/1), STFC capital grants ST/H008861/1 and ST/H00887X/1, and STFC DiRAC Operations grant ST/K00333X/1. DiRAC is part of the National E-Infrastructure.

\def\aj{{AJ}}                   
\def\araa{{ARA\&A}}             
\def\apj{{ApJ}}                 
\def\apjl{{ApJ}}                
\def\apjs{{ApJS}}               
\def\ao{{Appl.~Opt.}}           
\def\apss{{Ap\&SS}}             
\def\aap{{A\&A}}                
\def\aapr{{A\&A~Rev.}}          
\def\aaps{{A\&AS}}              
\def\azh{{AZh}}                 
\def\baas{{BAAS}}               
\def\jrasc{{JRASC}}             
\def\memras{{MmRAS}}            
\def\mnras{{MNRAS}}             
\def\pra{{Phys.~Rev.~A}}        
\def\prb{{Phys.~Rev.~B}}        
\def\prc{{Phys.~Rev.~C}}        
\def\prd{{Phys.~Rev.~D}}        
\def\pre{{Phys.~Rev.~E}}        
\def\prl{{Phys.~Rev.~Lett.}}    
\def\pasp{{PASP}}               
\def\pasj{{PASJ}}               
\def\qjras{{QJRAS}}             
\def\skytel{{S\&T}}             
\def\solphys{{Sol.~Phys.}}      
\def\sovast{{Soviet~Ast.}}      
\def\ssr{{Space~Sci.~Rev.}}     
\def\zap{{ZAp}}                 
\def\nat{{Nature}}              
\def\iaucirc{{IAU~Circ.}}       
\def\aplett{{Astrophys.~Lett.}} 
\def\apspr{{Astrophys.~Space~Phys.~Res.}}   
\def\bain{{Bull.~Astron.~Inst.~Netherlands}}    
\def\fcp{{Fund.~Cosmic~Phys.}}  
\def\gca{{Geochim.~Cosmochim.~Acta}}        
\def\grl{{Geophys.~Res.~Lett.}} 
\def\jcp{{J.~Chem.~Phys.}}      
\def\jgr{{J.~Geophys.~Res.}}    
\def\jqsrt{{J.~Quant.~Spec.~Radiat.~Transf.}}   
\def\memsai{{Mem.~Soc.~Astron.~Italiana}}   
\def\nphysa{{Nucl.~Phys.~A}}    
\def\physrep{{Phys.~Rep.}}      
\def\physscr{{Phys.~Scr}}       
\def\planss{{Planet.~Space~Sci.}}           
\def\procspie{{Proc.~SPIE}}     
\def\actaa{{Acta~Astronomica}}     
\def\pasa{{Publications of the ASA}}     
\def\na{{New Astronomy}}     
\def\icarus{{Icarus}}     

\let\astap=\aap
\let\apjlett=\apjl
\let\apjsupp=\apjs
\let\applopt=\ao
\let\mnrasl=\mnras

\bsp

\label{lastpage}

\end{document}